\documentclass[prx,twocolumn,english,superscriptaddress,floatfix,longbibliography,nofootinbib, aps, 10pt]{revtex4-2}
\usepackage[toc,page]{appendix}
\usepackage[utf8]{inputenc}
\pdfoutput=1

\usepackage{silence}
\usepackage{silence}
\usepackage{adjustbox}

\usepackage[utf8]{inputenc}
\DeclareUnicodeCharacter{2212}{-}
\DeclareUnicodeCharacter{00A0}{ }

\usepackage[T1]{fontenc}
\usepackage{lmodern}
\usepackage{setspace}
\usepackage[usenames, dvipsnames]{xcolor}
\usepackage{graphicx}
\usepackage{mathtools}
\usepackage{amssymb}
\usepackage{amsthm}
\usepackage{xspace}
\usepackage{booktabs}
\usepackage{tikz-cd}
\usepackage[backref=page, bookmarks=false]{hyperref}
\usepackage[capitalize, noabbrev]{cleveref}
\usepackage{slashed}

\usepackage[usenames, dvipsnames]{xcolor}
\usepackage{tikz}
\usepackage{spectralsequences}
\usepackage{subcaption}  

\newcommand{\mb}{\mathbb}

\newcommand{\Z}{\mathbb{Z}}

\newcommand{\ko}{\mathit{ko}}

\newcommand{\U}{\mathrm U}

\newsavebox{\pullback}
\sbox\pullback{%
\begin{tikzpicture}%
\draw (0,0) -- (1ex,0ex);%
\draw (1ex,0ex) -- (1ex,1ex);%
\end{tikzpicture}}

\newtheorem*{thm*}{Theorem}

\theoremstyle{definition}

\theoremstyle{remark}

\crefname{thm}{Theorem}{Theorems}
\crefname{lem}{Lemma}{Lemmas}
\crefname{cor}{Corollary}{Corollaries}
\crefname{prop}{Proposition}{Propositions}
\crefname{ex}{Exercise}{Exercises}
\crefname{exm}{Example}{Examples}
\crefname{defn}{Definition}{Definitions}
\crefname{claim}{Claim}{Claims}
\crefname{rem}{Remark}{Remarks}
\crefname{fct}{Fact}{Facts}
\crefname{note}{Note}{Notes}

\DeclarePairedDelimiter\paren{(}{)}

\makeatletter
	\let\oldparen\paren
	\def\paren{\@ifstar{\oldparen}{\oldparen*}}
\makeatother

\usepackage{microtype}
\usepackage{hypcap}

\hypersetup{
 colorlinks,
 linkcolor={red!50!black},
 citecolor={green!50!black},
 urlcolor={blue!80!black}
}

\newcommand{\NewThomSpectrum}[1]{\expandafter\newcommand\csname M#1\endcsname{\mathit{M#1}}}
\newcommand{\NewMTSpectrum}[1]{\expandafter\newcommand\csname MT#1\endcsname{\mathit{MT#1}}}
\newcommand{\BothThomSpectra}[1]{\NewThomSpectrum{#1}\NewMTSpectrum{#1}}
\BothThomSpectra{O}
\BothThomSpectra{SO}
\BothThomSpectra{Spin}
\BothThomSpectra{Pin}
\BothThomSpectra{U}
\BothThomSpectra{SU}

\usepackage{amstext} 
\usepackage{array}   
\newcolumntype{L}{>{$}l<{$}} 
\newcolumntype{C}{>{$}c<{$}} 

\newcommand{\Pin}{\mathrm{Pin}}
\newcommand{\Spin}{\mathrm{Spin}}

\newcommand{\SO}{\mathrm{SO}}

\newcommand{\SU}{\mathrm{SU}}

\newcommand{\Spinc}{\relax\ifmmode{\Spin^c}\else Spin\textsuperscript{$c$}\xspace\fi}
\newcommand{\spinc}{spin\textsuperscript{$c$}\xspace}

\newcommand{\Pinc}{\relax\ifmmode{\Pin^c}\else Pin\textsuperscript{$c$}\xspace\fi}

\newcommand{\Pinp}{\relax\ifmmode{\Pin^+}\else Pin\textsuperscript{$+$}\xspace\fi}

\newcommand{\Pinm}{\relax\ifmmode{\Pin^-}\else Pin\textsuperscript{$-$}\xspace\fi}

\usepackage{graphicx}
\usepackage{dcolumn}
\usepackage{bm}
\usepackage{bbm}
\usepackage{framed}
\usepackage{setspace}
\usepackage{float}
\usepackage{tcolorbox}
\usepackage{subfiles}

\def\<{\langle}
\def\>{\rangle}

\def\bZ{\mathbb{Z}}
\def\bR{\mathbb{R}}

\def\cO{\mathcal{O}}

\DeclareTextFontCommand{\df}{\bf}

\usepackage{amsthm}

\theoremstyle{definition}
\newtheorem{defin}{Definition}

\usepackage{tikz}
\usepackage{tikz-cd}
\usepackage{color}
\usepackage{xcolor}
\usepackage{enumitem}

\usetikzlibrary{arrows}

\def\[#1\]{%
  \begin{equation}\begin{gathered}#1\end{gathered}\end{equation}%
}

\date{\today}

\begin{document}

\title{\LARGE A Long Exact Sequence in Symmetry Breaking: \\ \large order parameter constraints, defect anomaly-matching, and higher Berry phases}

\author{Arun Debray}
\email{adebray@purdue.edu}
\affiliation{Department of Mathematics, Purdue University, 150 N University Street, West Lafayette, IN 47907, USA}

\author{Sanath K. Devalapurkar}
\email{sdevalapurkar@math.harvard.edu}
\affiliation{Harvard University Department of Mathematics,
1 Oxford Street,
Cambridge, MA 02138, USA}

\author{Cameron Krulewski}
\email{camkru@mit.edu}
\affiliation{Massachusetts Institute of Technology,
Department of Mathematics,
Simons Building, 
77 Massachusetts Avenue,
Cambridge, MA 02139, USA}

\author{Yu Leon Liu}
\email{yuleonliu@math.harvard.edu}
\affiliation{Harvard University Department of Mathematics,
1 Oxford Street,
Cambridge, MA 02138, USA}

\author{Natalia Pacheco-Tallaj}
\email{nataliap@mit.edu}
\affiliation{Massachusetts Institute of Technology,
Department of Mathematics,
Simons Building, 
77 Massachusetts Avenue,
Cambridge, MA 02139, USA}

\author{Ryan Thorngren}
\email{ryan.thorngren@physics.ucla.edu}
\affiliation{Mani L. Bhaumik Institute for Theoretical Physics, Department of Physics and Astronomy,
University of California, Los Angeles, CA 90095, USA}

\begin{abstract}
We study defects in symmetry breaking phases, such as domain walls, vortices, and hedgehogs. In particular, we focus on the localized gapless excitations which sometimes occur at the cores of these objects. These are topologically protected by an 't Hooft anomaly. We classify different symmetry breaking phases in terms of the anomalies of these defects, and relate them to the anomaly of the broken symmetry by an anomaly-matching formula. We also derive the obstruction to the existence of a symmetry breaking phase with a local defect. 
We obtain these results using a long exact sequence of groups of invertible field theories, which we call the ``symmetry breaking long exact sequence'' (SBLES). 
The mathematical backbone of the SBLES
is studied in a companion paper \cite{MathSmith}.
Our work further develops the theory of higher Berry phase and its bulk-boundary correspondence, and serves as a new computational tool for classifying symmetry protected topological phases.
 
\end{abstract}
\maketitle

\tableofcontents

\section{Introduction}

Over the past twenty years or so, there has been a revolution in the way we understand symmetries and anomalies of many-body quantum systems, both in the continuum and on the lattice, spurred by the discovery of topological insulators and other condensed matter systems exhibiting bulk-boundary correspondence, or anomaly in-flow. In this paper, we study phenomena associated with symmetry breaking at the surface of such phases, and in particular the gapless modes localized at domain walls, vortices, hedgehogs, and other defects in the order parameter, of a broad class that we define in this work. In particular, we provide a complete solution of the anomaly matching problem for such surface defects, relating their classification to that of the bulk phase.

For example, the surface of a 3d topological insulator famously supports a single Dirac cone, protected by charge conservation $U(1)$ and time reversal symmetry. When brought into contact with a superconductor (thought of as a $U(1)$ symmetry breaking state), even if the superconductor is a normal $s$-wave state, an exotic sort of superconductivity occurs at the interface by proximity effect \cite{Fu_2008,seiberg2016gapped}, characterized by Majorana zero modes at vortices. Other famous examples of localized gapless modes include chiral modes along domain walls \cite{jackiwrebbe} and axion strings \cite{callan1985anomalies,goldstonewilczek}.

It turns out that in many cases, the existence of localized gapless modes at such defects is guaranteed by anomaly matching, and holds even at strong coupling. An anomaly matching formula of this type was first provided in \cite{HKT19}, although it was noticed that 1. not all anomalies are consistent with local defects in a symmetry breaking phase and 2. even when it exists, the anomaly of the defect is not determined by the anomaly matching formula. Determining the constraints under which defect anomaly matching can be applied and its ambiguities were left as open problems.

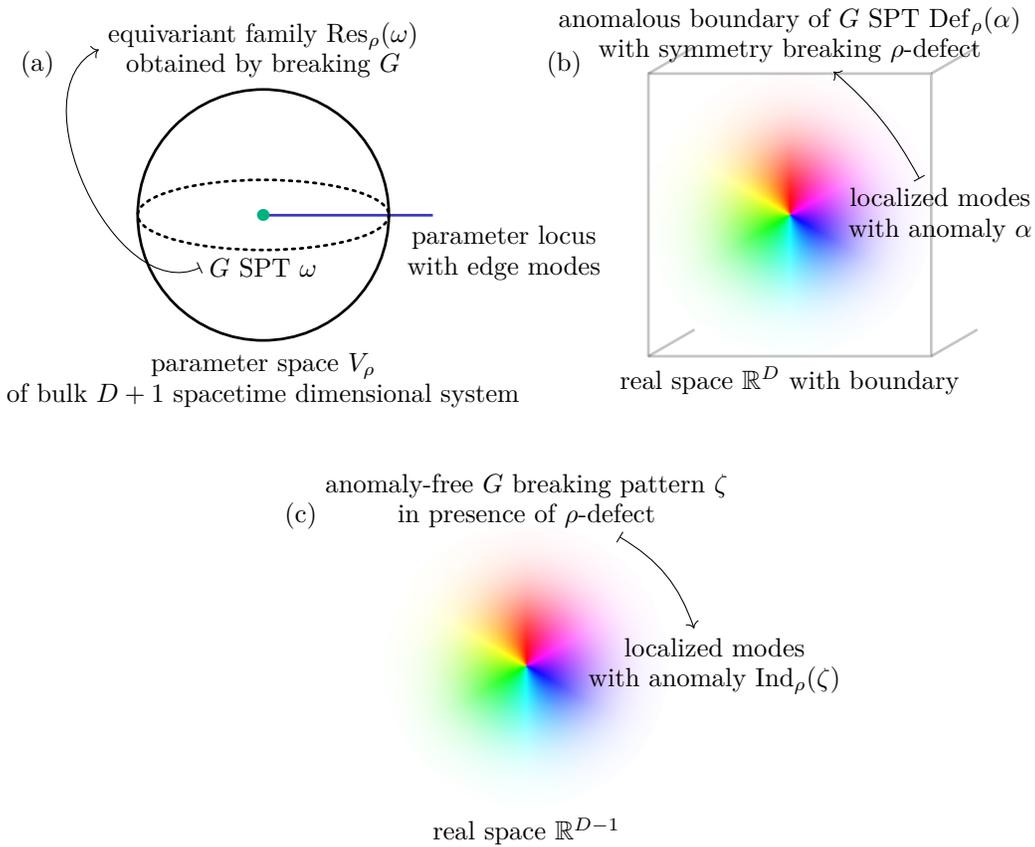
\begin{figure*}
    \centering
    \begin{tikzpicture}
    \node at (0,0) {\includegraphics[width=5cm]{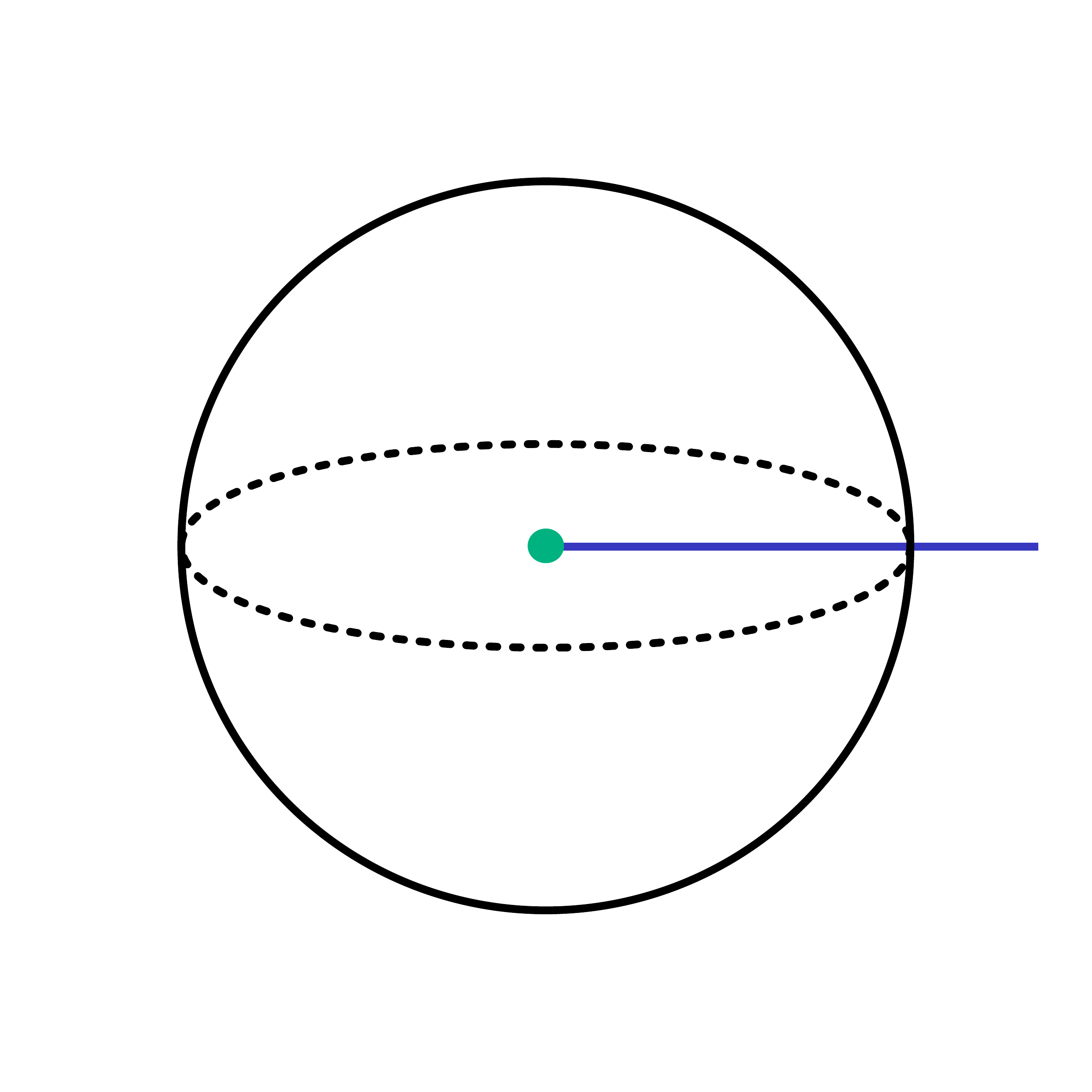}};
    \node at (7,0) {\includegraphics[width=5cm]{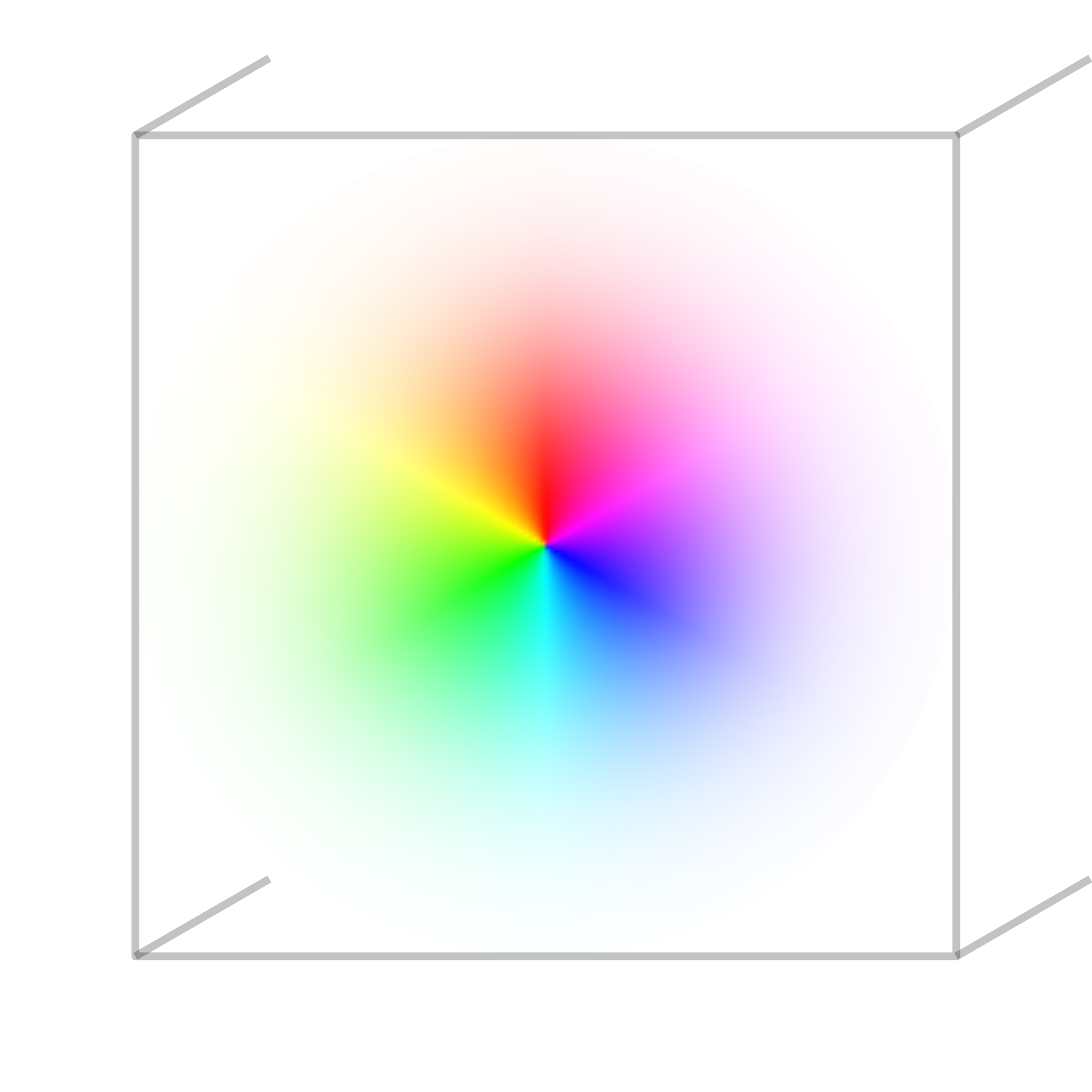}};
    \node at (-3,2) {(a)};
    \node at (4,2) {(b)};
    \node at (0.5,-4) {(c)};
    \node (2) at (0,2.2) [align=center]{equivariant family ${\rm Res}_\rho(\omega)$ \\ obtained by breaking $G$};
    \node at (0,-2.2) [align=center]{parameter space $V_\rho$ \\ of bulk $D+1$ spacetime dimensional system};
    \node (1) at (0,-.7) {$G$ SPT $\omega$};
    \path[|->] (1) edge[bend right=-90] node [left] {} (2);
    \node at (3.2,-0.5) [align=center]{parameter locus \\ with edge modes};
    \node at (7,-2.2) {real space $\bR^D$ with boundary};
    \node (4) at (7,2.4) [align=center]{anomalous boundary of $G$ SPT ${\rm Def}_\rho(\alpha)$ \\ with symmetry breaking $\rho$-defect};
    \node (3) at (9,0) [align=center]{localized modes \\ with anomaly $\alpha$};
    \path[|->] (3) edge[bend right=10] node [left] {} (4);
    \node at (3.5,-6) {\includegraphics[width=5cm]{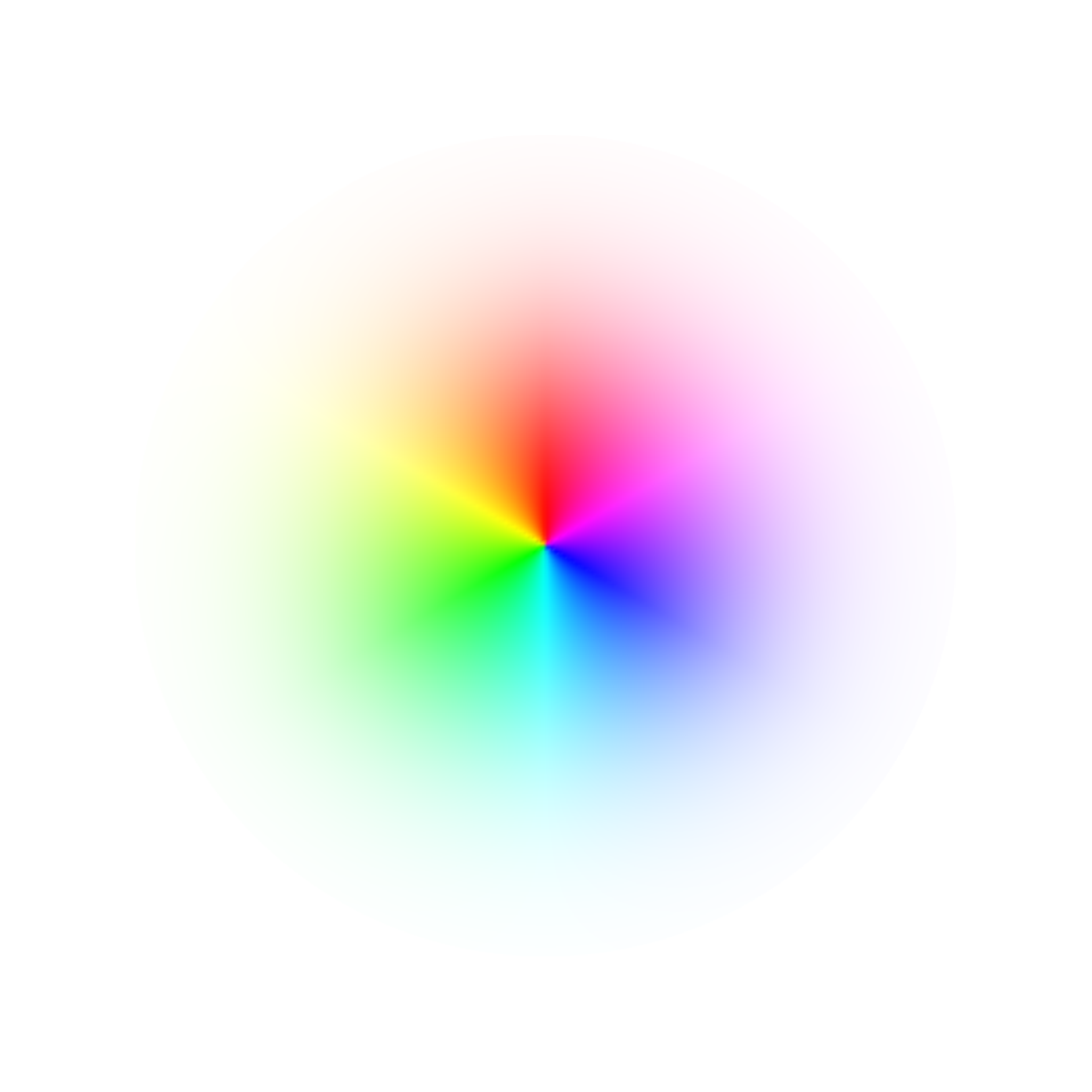}};
    \node (6) at (6,-6) [align=center]{localized modes \\ with anomaly ${\rm Ind}_\rho(\zeta)$};
    \node at (3.5,-8.2) {real space $\bR^{D-1}$};
    \node (5) at (3.5,-3.8) [align=center] {anomaly-free $G$ breaking pattern $\zeta$ \\ in presence of $\rho$-defect};
    \path[|->] (5) edge[bend right=-20] node [left] {} (6);
    \end{tikzpicture}
    \caption{\textbf{The three anomaly-matching maps:} \textbf{(a)} (\cref{subsecfamilyanomaly}) applying a symmetry breaking field transforming in the representation $\rho$ to the $G$ SPT $\omega$ produces a $G$-equivariant invertible family ${\rm Res}_\rho(\omega)$ on the unit sphere $S(\rho)$. When this anomaly-free $G$ breaking pattern is topologically nontrivial, there is a parameter locus where the boundary gap closes (a diabolical locus in the sense of \cite{higherberry}). This locus begins at the origin, where we have $G$ symmetry and protected SPT edge modes, but even though $G$ is broken it extends to infinity. This is the obstruction to a local $\rho$-defect on the boundary, and we call it the residual family anomaly. \textbf{(b)} (\cref{subsecdefectanomap}) When an SPT satisfies ${\rm Res}_\rho(\omega) = 0$, there is a local $\rho$-defect on the boundary, a class of defect including domain walls, vortices, hedgehogs, etc, which may host localized modes with anomaly $\alpha$. The defect anomaly map (aka the Smith homomorphism of \cite{HKT19}) reconstructs from $\alpha$ the bulk SPT as $\omega = {\rm Def}_\rho(\alpha)$. \textbf{(c)} (\cref{subsecindexmap}) The defect anomaly map can reconstruct the boundary anomaly but it cannot generally be inverted to give the anomaly of the defect. Indeed, even in an anomaly-free $G$ equivariant invertible family $\zeta$, we can have a $\rho$-defect with localized anomalous modes. The index map computes their anomaly as $\alpha = {\rm Ind}_\rho(\zeta)$. This gives the ambiguity in the boundary $\rho$-defect in item (b) and a generalization of the Callias index theorem to interacting systems. In turn, families of the form ${\rm Res}_\rho(\omega)$ (as in (a)) are precisely those with trivial index maps, completing the circle (\cref{subseccompletingthecircle}).}
    \label{fig:summary}
\end{figure*}

In this work, we devise a general theory of defect anomaly matching, in terms of a mathematical object known as a long exact sequence, which captures both the obstruction to the existence of a symmetry breaking phase with a local defect and the classification of such phases. The results are summarized in \cref{fig:summary}, with details to be explained later.

The physical input relies on the recent concept of higher Berry phase and its associated bulk-boundary correspondence \cite{higherberry,Cordovaberry,antonlevberry,ashvinberry} which we also further develop. In particular, we formulate an interacting version of the Callias index theorem \cite{callias1978axial,bott1978some} which we believe will have further applications.

As a computational tool, our long exact sequence turns out to be remarkably convenient. Different symmetry breaking patterns can be combined to calculate the classification of anomalies for a given symmetry group and dimension, often avoiding difficult spectral sequence calculations. 
Other papers using this or closely related techniques to do computations include~\cite{HS13, Deb23, DDHM22, DL23, DYY}.

The outline of the paper is as follows:

In \cref{subsecanomalyintro} we review the description and classification of 't Hooft anomalies in terms of invertible field theories, including some more recent perspectives and family anomalies.

\cref{SBLES_section} contains the description of the symmetry breaking long exact sequence (SBLES) and our physical results. The SBLES consists of three anomaly-matching formulas/maps: (\cref{subsecfamilyanomaly}) the residual family anomaly which persists after explicitly breaking the global symmetry and which provides the obstruction to a local defect in the order parameter; (\cref{subsecdefectanomap}) the defect anomaly map which reconstructs the bulk anomaly from the anomaly of the local defect, when it exists; and (\cref{subsecindexmap}) the index map which describes the anomaly of a defect in an invertible family and which determines the ambiguity of the defect anomaly in terms of the classification of topologically distinct symmetry breaking patterns of one lower dimension, thus coming in full circle. We discuss each of these in turn with several examples, before putting them all together in a long exact sequence in \cref{subseccompletingthecircle}.

We discuss extended examples in \cref{sec:examples-of-SBLES} and provide the analogous long exact sequence in group cohomology in \cref{sec:group_cohomology}.

\noindent\textbf{Acknowledgements.} We would like to thank
Adrian Clough,
Dan Freed,
Mike Hopkins,
Theo Johnson-Freyd,
Justin Kulp,
Miguel Montero,
David Reutter,
Luuk Stehouwer,
Weicheng Ye, and
Matthew Yu
for helpful discussions which improved our paper.

Part of this project was completed while AD, CK, and YLL visited the Perimeter Institute for Theoretical Physics for the 2022 workshop on Global Categorical Symmetries; research at Perimeter is supported by the Government of Canada through Industry Canada and by the Province of Ontario through the Ministry of Research \& Innovation. The workshop on Global Categorical Symmetries was supported by the Simons Collaboration on Global Categorical Symmetries. We thank both Perimeter and the Simons Collaboration for Global Categorical Symmetries for their hospitality.
YLL is supported by the Simons Collaboration on Global Categorical Symmetries.
CK and NPT are supported by NSF DGE-2141064 and SKD is supported by NSF DGE-2140743.

\section{Anomalies and invertible field theories}\label{subsecanomalyintro}

\subsection{$G$-anomalies}
\label{sss:g_anom}

An 't Hooft anomaly for a global symmetry $G$ (or just $G$-anomaly) can be roughly defined as an obstruction to gauging $G$. This typically appears in some gauge non-invariance when we couple our theory to a background gauge field $A$. Let us write the partition function on a spacetime $X^D$ with this background as $Z(X^D,A)$. Under a gauge transformation $A \mapsto A^g$, we may have
\[Z(X^D,A^g) = e^{i \alpha(X,A,g)} Z(X^D,A),\]
where $e^{i\alpha(X^D,A,g)}$ is some phase factor which signals that $Z(X^D,A)$ is not gauge invariant and there may be an anomaly. More precisely, since $Z(X^D,A)$ is only defined up to local counter-terms, $\alpha(X^D,A,g)$ is only defined up to variations of local counterterms, and if $\alpha$ cannot be cancelled this way, there is a $G$-anomaly.

Under mild assumptions about $\alpha(X^D,A,g)$ (see Section 5 of \cite{Thorngren_2021}), and in all known cases, there is a local counterterm $e^{i\omega(Y^{D+1},A)}$ defined in \emph{one greater dimension} so that if $\partial Y^{D+1} = X^D$, then
\[e^{i \omega(Y^{D+1},A^g) - i \omega(Y^{D+1},A)} = e^{i\alpha(X^D,A,g)}.\]
This is called anomaly in-flow, since for continuous $G$ it can be interpreted as missing boundary charge flowing
into the bulk, and allows us to relate $G$-anomalies in $D$-dimensions to local counterterms $e^{i
\omega(Y^{D+1},A)}$ in $D+1$ dimensions. The phase factor $e^{i \omega(Y^{D+1},A)}$ itself is the (phase of the)
partition function of a particularly simple type of $D+1$-dimensional theory known as a $G$-symmetric invertible
field theory. These theories are so named because if we take stacks of such theories (which multiplies their
partition functions), each theory has an inverse with which it stacks to the trivial theory.

A famous example is the chiral anomaly in 1+1d. We have a theory of a free Dirac fermion with independently conserved left-movers and right-movers, corresponding to a symmetry group $G = U(1)_L \times U(1)_R$ with generators $L$ and $R$. If we turn on background gauge fields $A_L$ and $A_R$ each with $2\pi$ magnetic flux through spacetime $X$, then there will be fermion zero modes which must be subtracted from the path integral measure, leading to an imbalance of ``axial'' $L-R$ charge and a nontrivial gauge variation of $Z_{\rm Dirac}(X,A_L,A_R)$. This variation is equivalent to the boundary variation of the 2+1d Chern-Simons term \cite{witten1989quantum,Wen_1995}
\[\label{eqnchiralanomchernsimons}\omega(Y^3,A) = \frac{1}{4\pi} \int_{Y^3} A_L dA_L - A_R dA_R.\]
We can think of the Dirac fermion as living at the boundary of a theory with this partition function. If we make symmetric deformations of the Dirac fermion, such as adding Luttinger interactions, the bulk cannot be affected even at strong coupling, and hence the anomaly does not change, since it is determined by the bulk. This property, known as anomaly matching, makes anomalies very useful for studying phase diagrams of theories and renormalization group flows.

Because of this bulk-boundary correspondence, we can study anomalies by studying the invertible field theory in the
bulk. Invertible field theories are rather simple as physical theories, having just a single state in their Hilbert
space associated to each closed manifold. However, as mathematical objects they are quite rich, and are expected to
form an object called a loop spectrum. This roughly means that a family of invertible field theories in $D$
dimensions parametrized by $S^1$ (i.e. $S^1$-family) is equivalent to an invertible field theory in $D-1$
dimensions. The equivalence is via a ``Thouless pump'', where the $D-1$-dimensional invertible field theory gets
``pumped'' to the boundary when we go adiabatically around the $S^1$-family in $D$ dimensions
\cite{kiteavtalk,Xiong_2018,Gaiotto_2019}. The main technical result of our work, the symmetry breaking long exact sequence 
(see \cref{SBLES_section}), can be derived
from the loop spectrum property. However, for concreteness and ease of calculation, we will demonstrate our physics
results using a stronger conjectural description of these theories via cobordism theory, which we presently describe.

The SPT-cobordism conjecture \cite{kapustin2014symmetry,KTTW15,FH16} is that the loop spectrum associated to invertible field theory is the so-called Anderson dual of the Thom spectrum, which is related to the cobordism theory of manifolds. We will describe here the basic physics content of this conjecture. First we must define a cobordism. A cobordism between two manifolds $M_1$ and $M_2$ is a third manifold $N$ with $\partial N = M_1 \cup M_2$. Note we can define cobordisms for manifolds $M_1,M_2$ with structures like $G$ gauge fields by asking that the structure extends to the cobordism $N$. A cobordism invariant is something which is additive under disjoint union, and equal for all manifolds related by cobordism. The second condition can be stated that if $M = \partial N$, all cobordism invariants must be trivial for $M$, since $N$ gives a cobordism between $M$ and the empty manifold.

The SPT-cobordism conjecture roughly means that $e^{i\omega(Y^{D+1},A)}$ behaves like the holonomy of a $D+1$-form connection integrated over $Y^{D+1}$ \cite{yamashita2023invertible,Yam21,YY21}. In particular, there are ``Chern numbers'' associated with this connection, which are integer-valued cobordism invariants of closed $D+2$ manifolds (equipped with a $G$ gauge field and any other relevant structure). We can think of this integer as the winding number of $e^{i\omega(Y^{D+1},A)}$ evaluated along slices of the $D+2$ manifold (compare \cite{higherberry}). Deformation equivalence classes (meaning continuous deformation within the space of invertible field theories, i.e. $\pi_0$ of this space) of invertible field theories are believed to be classified by these invariants.

In practice, this means $e^{i\omega(Y^{D+1},A)}$ can be written as a product of two terms: $(1)$ a Chern-Simons invariant evaluated on $Y^{D+1}$, which is itself associated with an integer cobordism invariant in $D+2$ dimensions (now two more than the anomalous theory!), e.g.
\[\frac{1}{8\pi^2}\int_{W^4} dA_L dA_L - dA_R dA_R\]
is associated with \eqref{eqnchiralanomchernsimons}; and $(2)$ a $U(1)$-valued cobordism invariant in $D+1$-dimensions evaluated on $Y$, which typically consists of torsion pieces (valued in a finite subgroup of $U(1)$) and theta angles (which are not fixed under deformations). Note that we also equip $Y$ with a metric, so that (1) can also include gravitational Chern-Simons terms.

\subsection{Family anomalies}

Besides $G$-anomalies, we are also interested in \emph{family anomalies}, a relatively new concept which has appeared in the study of theories with a parameter space \cite{crystallineSPT,KSTX17,toptermsandphasesofsigmas,Cordovaberry,antonlevberry,higherberry,ashvinberry}. Suppose we have a theory depending on a parameter space $M$. We can couple the theory to a background field $\phi(x) \in M$ for these parameters and consider $Z(X^D,\phi)$. It may be that $Z(X^D,\phi)$ cannot be consistently defined over the space of background fields, and instead behaves like a section of a line bundle. This is analogous to how a quantum mechanical system with a nontrivial Berry number cannot have a globally defined ground state.

In practical terms, the family anomaly for a collection of local operators $\cO_1,\ldots,\cO_n$ is an obstruction to choosing a local Hamiltonian $H_0$ such that\footnote{Note that, just as $G$-anomalies are only a property of the $G$ action on the microscopic degrees of freedom, not of the dynamics, so too does the family anomaly only depend on the operators we couple to.}
\[H(c_1,\ldots,c_n) = H_0 + \sum_j c_j \int d^d x \cO_j(x)\]
has a gapped, nondegenerate ground state for all $\sum_j |c_j|^2 > C$, for some $C$. This makes family anomalies especially useful for studying phase diagrams.

Family anomalies in $D$ dimensions are associated with boundaries of theories in $D+1$-dimensions with a
\emph{higher} Berry number \cite{higherberry}. We may consider these higher dimensional theories to be invertible
field theories for spacetimes equipped with the parameter field $\phi$. The boundary partition function
$Z(X^D,\phi)$ is then considered a vector in the (1d) state space of this theory (à la relative QFT \cite{freed2014relative}). Considered this way, family anomalies are actually a generalization of $G$-anomalies, since we may take $M = BG$.

It is interesting to combine family anomalies and $G$-anomalies, especially when there is \emph{explicit} symmetry breaking. In the simple case with no explicit symmetry breaking, meaning for every value of the parameters $M$ we have $G$-symmetry, we call this a $G$-symmetric family. More interesting is the case of a \emph{$G$-equivariant family}, where $G$ acts nontrivially on $M$, such that if $m \in M$ is fixed by some subgroup $G_m < G$, the theory at that parameter value has $G_m$ symmetry. Other elements $g \in G$ map states and observables at $m$ to those at $g \cdot m$.

When we have a $G$-equivariant family and we turn on a background gauge field $A$, the parameter field $\phi$ can no longer be a globally defined map to $M$. Instead, we can think of it as having boundary conditions set by the transition functions of the $G$ gauge bundle $P \to X$ \cite{crystallineSPT}. More precisely, we can define the associated $M$-bundle $P \times_G M \coloneqq P \times M/G^{\rm diagonal} \to X$ by the action of $G$ on $M$, and define $\phi$ to be a section of this bundle. If it is possible to couple to such a background, we call the family anomaly-free. Note that not all $G$-equivariant families of invertible field theories are anomaly-free, but the ones characterizing family anomalies always are.

$G$-equivariant family anomalies in $D$ dimensions are thus classified by invertible field theories in $D+1$ dimensions for spacetimes equipped with a $G$-gauge field $A$ and a parameter field $\phi$, which is a section of the associated $M$-bundle \cite{higherberry}. These are described in cobordism theory as above, where we ask that $\phi$ also extends to the cobordism.

We give an example of an equivariant family anomaly occurring at the 0+1d boundary of a 1+1d system. The 1+1d system is constructed beginning with the free Dirac fermion $\psi$ we considered above. We will have a parameter space $M = S^1$ with a $2\pi$-periodic coordinate $\theta$, which parametrizes the mass deformation $\cos \theta \bar \psi \psi + i \sin \theta \bar \psi \gamma^c \psi$, where $\gamma^c = \gamma^0 \gamma^1$. This breaks $G = U(1)_L \times U(1)_R$ to the diagonal ``vector'' subgroup $U(1)_V$ with generator $V = L + R$. The ``axial'' subgroup with generator $A = L - R$ is broken down to $\bZ/2$, and acts on $M$ as a rotation $\theta \mapsto \theta + 2\alpha$, where $\alpha$ is the angle of the axial rotation. Thus the family is not $G$-symmetric, but it is $G$-equivariant since an axial rotation just acts on the parameter $\theta$.

Let us promote the parameter to a background field $\theta = \phi(x,t)$. If we compute the vector current in this model, as a result of the chiral anomaly, we will find a contribution proportional to $\partial_t \phi$, which results in a ``Thouless pump'': adiabatically taking the parameter around a $2\pi$ cycle causes a single $U(1)_V$ charge to be transported across the system \cite{thouless1983quantization}. This results in a family anomaly at the boundary, since we cannot define the $U(1)_V$ charge there, consistently over the parameter space. As a result of this anomaly, given any $U(1)_V$-symmetric boundary condition, there will be some value of $\theta$ where the boundary gap closes, with two states of different $U(1)_V$ charge crossing in energy \cite{higherberry}. This generalizes the famous Jackiw-Rebbi domain wall zero mode \cite{jackiwrebbe}.

We can derive the bulk topological term associated with this family anomaly
\[\frac{1}{2\pi} \int_{Y^2} \phi dA_V.\]
Indeed, by varying $A_V$, we find the contribution to the vector current $\partial_t \phi$ which characterizes the Thouless pump. It also defines the $(1+1)$-dimensional invertible field theory which characterizes the boundary family anomaly. Broadly speaking, the boundary family anomaly in explicit symmetry breaking situations like this one can be derived directly from the bulk anomaly. We will spend much of the paper explaining how this works in general, and also return to this and related examples of massive free fermions.

\subsection{Twisted tangent structures}\label{physics_tangential}

We will also need certain tangent structures on our spacetime manifolds, which are required to consistently define the microscopic degrees of freedom of the theory. For example, we may need a metric and an orientation to define basic kinetic terms, and in this paper we will always ask for these structures. In fermionic theories, we will further ask for a Spin structure, which is needed in the UV to define consistent boundary conditions for fermions. The anomaly typically depends on these choices, and we will need the invertible field theory in one more dimension to be equipped with these data as well.

The presence of the background $G$ gauge field can ``twist'' these structures. For example, if $G = U(1)$ and we have a spin-charge relation, with all (fermionic) bosonic operators having (half) integer charge, respectively, then fermions can be defined using a Spin$^c$ structure \cite{seiberg2016gapped}. This is slightly weaker than a Spin structure, but requires some compatibility between the background gauge field $A$ and the tangent bundle of spacetime. In particular, we have
\[\oint_\Sigma \frac{1}{2\pi} dA + \frac{1}{2} w_2(TX) \in \bZ\]
for all closed surfaces $\Sigma$, where $w_2(TX)$ is the 2nd Stiefel-Whitney class of the tangent bundle $TX$, i.e. the obstruction to choosing a Spin structure on $TX$. We can think of the spin-charge relation in general as defining a central extension of $G$ by fermion parity $\bZ/2^F$
\[\bZ/2^F \to G_F \to G,\]
where $G_F$ has linear (as opposed to projective) representations on fermionic operators. We can classify such central extensions by a class in $H^2(G,\bZ/2)$.

The other sort of twist which is important occurs with spacetime-orientation-reversing symmetries. For example, suppose $G = \bZ/2$ acts as a time reversal symmetry. If $\gamma \subset X$ is a closed loop in spacetime around which the background gauge field $A$ has nontrivial holonomy
\[\int_\gamma A = 1 \mod 2,\]
then it will be impossible to choose a consistent orientation of $X$ around this loop, since we reverse the direction of time as we go around it\footnote{We are working in a Euclidean picture, so there is no special time coordinate.} \cite{kapustin2014symmetry}. Thus we are forced to consider non-orientable spacetimes. Likewise, to define our theory on such manifolds, we \emph{must} have nontrivial holonomy for $A$ along orientation reversing loops such as $\gamma$. We can phrase this compatibility condition between $A$ and the tangent bundle as follows:
\[\int_\gamma A = \int_\gamma w_1(TX) \mod 2,\]
for all closed loops $\gamma$, where $w_1(TX)$ is the 1st Stiefel-Whitney class of $TX$. We can classify the spacetime-orientation-reversing elements of $G$ as a homomorphism $G \to \bZ/2$, or equivalently a class in $H^1(G,\bZ/2)$.

A convenient way to encode both these data is to say that we have an orientation, Spin structure, etc. not on the tangent bundle $TX$ of spacetime, but on the direct sum $TX \oplus A^* \eta$, where $A^*\eta$ is a vector bundle associated to the $G$ gauge bundle by some $\bR$-linear $G$ representation $\eta$. The 1st and 2nd Stiefel-Whitney classes of $\eta$, considered as a vector bundle over the classifying space $BG$, define the twist classes $w_1(\eta) \in H^1(G,\bZ/2)$ and $w_2(\eta) \in H^2(G,\bZ/2)$ we considered above. Physically, we can think of $\eta$ as the representation of fermion bilinears in the theory \cite{KTTW15}, although our classification will only depend on the classes $w_1(\eta)$ and $w_2(\eta)$.

For example, suppose we study $G = \bZ/2$ global symmetry. $\bZ/2$ has a single nontrivial irrep, the sign representation $\sigma$. Let us take $\eta = n\sigma$, meaning a sum of $n$ copies of the sign representation. We find a four-fold periodic structure
\begin{itemize}
\item $n = 0 \mod 4$: ordinary $\bZ/2$ symmetry $U$ with $U^2 = 1$, corresponding to a separate Spin structure on $TX$ and a $\bZ/2$ gauge field.
\item $n = 1 \mod 4$: spacetime-orientation-reversing $\bZ/2$ symmetry $T$ with $T^2 = 1$, corresponding to a Pin$^-$ structure on $TX$ (see \cite{KT90} for an introduction to these structures).
\item $n = 2 \mod 4$: ordinary $\bZ/2$ symmetry $U$ with $U^2 = (-1)^F$, corresponding to a Spin$^c$ structure on $TX$ where the structure group of the determinant line is reduced from $U(1)$ to $\bZ/2$.
\item $n = 3 \mod 4$: spacetime-orientation-reversing $\bZ/2$ symmetry $T$ with $T^2 = (-1)^F$, corresponding to a Pin$^+$ structure on $TX$
\end{itemize}
This periodic structure is reflected in the repeated reduction of symmetry to the $\bZ/2$ domain wall \cite{HKT19}.

\subsection{The group of invertible field theories}\label{physics_TQFTs}

With all the data in hand, we are finally ready to define our object of interest:

\begin{defin}\label{defphysanom} Let $G$ be a group acting on a space $M$ (the parameter space), $s$ a tangent structure (usually an orientation aka SO structure in the case of bosonic theories or a Spin structure in the case of fermionic theories), $\eta$ a representation of $G$. We define $\Omega^n_{G,s,\eta}(M)$ to be the abelian group of deformation classes of invertible field theories defined for $n$-dimensional spacetimes $X$ equipped with a $G$-gauge field $A$, an $s$-structure on $TX \oplus A^*\eta$, a section $\phi$ of the $M$-bundle over $X$ associated with the gauge bundle of $A$, and a metric.
\end{defin}

Note that the group structure on invertible field theories corresponds to ``stacking'' of physical systems. That is, if we have two $D$-dimensional systems each with $G$ symmetry and parameter space $M$ depending on the same sort of tangent structure, then we can combine the two systems, initially decoupled, which will have $G \times G$ symmetry, a parameter space $M \times M$, and two tangent structures of the same kind. We want to preserve the diagonal $G < G \times G$, tune the parameters in tandem over the diagonal parameter space $M \hookrightarrow M \times M$, and couple to the same tangent structure in each ``layer''. Then we will have produced a third system in the same symmetry/parameter space/tangent structure class. We can do the same for the invertible field theories which determines the anomalies of each theory, and by definition the anomaly of the third system will be the sum of those two in the group structure thereof.

\section{The symmetry breaking long exact sequence}\label{SBLES_section}

In this section, we will outline our main result, summarized in \cref{fig:summary}, which is that three important maps (two of them new) in the theory of anomaly matching fit together into a ``long exact sequence.'' For us, a symmetry breaking pattern is described by a group $G$ and a real (orthogonal) representation $\rho$ of $G$ under which symmetry-breaking operators transform, for example in adding explicit symmetry-breaking terms to a Hamiltonian. Associated to this data, the symmetry breaking long exact sequence (SBLES) is

\begin{widetext}
\begin{center}
\begin{tcolorbox}
    \begin{tikzcd}[column sep=large]
  \hspace{.72in}  \cdots \to \Omega_{G,s,\eta}^D(S(\rho)) \arrow{r}{\rm{Ind}_\rho}
    & \Omega_{G,s,\eta+\rho}^{D+1-k} \arrow{r}{\rm{Def}_\rho} & \Omega_{G,s,\eta}^{D+1} \arrow{r}{{\rm Res}_\rho} & \Omega_{G,s,\eta}^{D+1}(S(\rho)) \to \cdots
    \end{tikzcd}
    \end{tcolorbox}
\end{center}
\end{widetext}
By anomaly in-flow, we can look at this long exact sequence either from the $D+1$-dimensional point of view of the invertible field theories, or from the $D$-dimensional point of view of the anomalous theories. From the latter point of view, the players are
\begin{itemize}
    \item $\Omega_{G,s,\eta}^{D+1}$: Anomalies of $G$-symmetric theories in $D$ spacetime dimensions, of type $s$ (bosonic or fermionic), and twist $\eta$.
    \item $\Omega_{G,s,\eta}^{D+1}(S(\rho))$: Anomalies of $G$-equivariant families of theories, parametrized by the unit sphere $S(\rho) \cong S^{k-1}$ in the representation $\rho$ (which has dimension $k$), and twist $\eta$.
\end{itemize}
See \cref{subsecanomalyintro} for a review of these groups. Meanwhile, the maps are
\begin{itemize}
    \item ${\rm Res}_\rho$: Measures the residual family anomaly of the $D$-dimensional theory after breaking the symmetry by an operator transforming in the representation $\rho$. (\cref{subsecfamilyanomaly})
    \item ${\rm Def}_\rho$: Describes the reconstruction of the bulk anomaly from the anomaly on a certain defect associated with this symmetry breaking, such as a domain wall. (\cref{subsecdefectanomap})
    \item ${\rm Ind}_\rho$: Encodes a generalized index theorem which associates an anomalous defect to a certain winding configuration in the space of symmetry-broken states. (\cref{subsecindexmap})
\end{itemize}
Each map has the property that its image is the kernel of the map following it. This is what makes it a ``long exact sequence.'' For example, those anomalies which have no residual family anomaly, and so live in the kernel of ${\rm Res}_\rho$, are precisely those which can be associated with a special defect, whose anomaly recovers the original anomaly by the map ${\rm Def}_\rho$. We show some long subsequences of the whole structure in \cref{subseccompletingthecircle}.

In this section, we will give physical definitions of each of these maps, arguments for the exactness of the sequence, and many examples of dynamical consequences of these maps. In later sections, we will give
longer examples. We offer mathematically precise definitions in our companion paper \cite{MathSmith}.

\subsection{Residual family anomalies}\label{subsecfamilyanomaly}

If we have a theory with a $G$-symmetry and an ('t Hooft) anomaly, there is no $G$-symmetric deformation of the theory to a nondegenerate, gapped phase. However, in the absence of gravitational anomalies, we can always nondegenerately gap the theory by breaking the symmetry, so long as we reduce to an anomaly-free subgroup $H < G$ (possibly trivial).

A more refined question is, if we have a family of symmetry breaking parameters transforming in a representation $\rho$ of $G$, when can we nondegenerately gap the theory for all large-enough values of the symmetry breaking parameters?

\begin{defin}

A theory is \textbf{$\rho$-(nondegenerately)-gappable} if there exists an operator transforming in the representation $\rho$, such that for all large enough perturbations by this operator (referred to as the symmetry breaking field), the theory has a (nondegenerate) gapped ground state. Equivalently, the ground state for all large enough symmetry breaking fields is \textbf{uniformly (nondegenerately) gapped}, meaning there is a uniform lower bound on the energy gap about the ground states (and further the ground state is unique). For this paper, ``nondegenerately'' will always be implied. This condition is equivalent to the existence of a local ``$\rho$-defect'', defined in \cref{subsecdefectanomap} below.
\end{defin}

It turns out that depending on the $G$ anomaly and $\rho$, a theory may not be $\rho$-gappable. The simplest such obstruction occurs when some unbroken symmetry $H$ is still anomalous, but we will derive the general obstruction. We find there are more subtle obstructions, which can exist even when all unbroken symmetries are anomaly free, and which are related to parameter space anomalies and higher Berry phases \cite{Cordovaberry,higherberry,antonlevberry,ashvinberry,toptermsandphasesofsigmas}.

The general obstruction can be derived by anomaly in-flow, as follows. We can start by thinking of our anomalous system as living at the boundary of a $D+1$-dimensional $G$-SPT, i.e. a $G$-symmetric invertible theory, (the equivalence class of) which we may use to label the 't Hooft anomaly. Let $V_\rho$ be the real vector space associated to $\rho$. For each value of the symmetry breaking field $v \in V_\rho$, we can extend the symmetry breaking into the SPT bulk.

This defines a $G$-equivariant family of $D+1$-dimensional invertible theories over $S(\rho)$, a $k-1$-sphere of large radius $S(\rho) \subset V_\rho$, with our original anomalous theory with symmetry breaking field defining a $G$-equivariant family of boundary conditions. 
The deformation class of the bulk defines a (linear) map
\[{\rm Res}_\rho\colon\Omega_{G,s,\eta}^{D+1}  \xrightarrow{} \Omega_{G,s,\eta}^{D+1}(S(\rho)).\label{eqnssbanomaly}\]

We call this map the \textbf{residual family anomaly}, since it turns out to be the obstruction to $\rho$-gappability. Indeed, if our $D$-dimensional theory is $\rho$-gappable and has anomaly $\omega \in \Omega_{G,s,\eta}^{D+1}$, then we must have ${\rm Res}_\rho (\omega) = 0$, since this would give us a uniformly gapped, $G$-equivariant family of boundary conditions for the $D+1$-dimensional invertible family with invariant ${\rm Res}_\rho (\omega)$, which is not possible if ${\rm Res}_\rho (\omega) \neq 0$, by the bulk-boundary correspondence for families \cite{higherberry}. Conversely, ${\rm Res}_\rho (\omega)$ is very likely the only obstruction to $\rho$-gappability, as we will argue below.

Recall that using the SPT-cobordism conjecture of \cref{subsecanomalyintro}, we can describe the anomaly $\omega \in \Omega_{G,s,\eta}^{D+1}$ as a function $\omega(X,A) \in U(1)$ on pairs of a spacetime $D+1$-manifold $X$ and a background gauge field $A$. Meanwhile ${\rm Res}_\rho (\omega) \in \Omega_{G,s,\eta}^{D+1}(S(\rho))$ can be described as a function $({\rm Res}_\rho (\omega))(X,A,\phi)$ on triples $(X,A,\phi)$ further consisting of a section $\phi$ as above. We can define this function by evaluating $\omega$ on just $(X,A)$, simply discarding $\phi$, giving
\[({\rm Res}_\rho (\omega))(X,A,\phi) \coloneqq \omega(X,A)\]

This residual family anomaly generalizes the anomaly of the unbroken symmetry. Indeed, consider the theory at some fixed $v\in S(\rho)$. This theory may have a residual anomaly for the unbroken subgroup $G_v < G$, which prevents us from gapping it without breaking $G_v$. The residual anomaly is thus also an obstruction to $\rho$-gappability. In fact, for each $v$, there is a map (pullback along the inclusion of $v$ in $S(\rho)$)
\begin{equation}\label{eqnpullbackfamily}
    \begin{aligned}
        v^*\colon\Omega_{G,s,\eta}^{D+1}(S(\rho)) &\to \Omega_{G_v,s,\eta}^{D+1} \\
        (v^* \omega)(X,A_v) &\coloneqq \omega(X,A_v,\phi = v),
    \end{aligned}
\end{equation}
such that the image of the $G$ anomaly under $v^* \circ {\rm Res}_\rho$ is the residual $G_v$ anomaly. So the residual family anomaly cannot vanish unless the residual anomaly also vanishes for each $v \in S(\rho)$. However, the examples below in Sections \ref{subsubsec2dMajoranaresidualanomaly} and \ref{subsubsecadjointqcdexample} demonstrate that even if all residual anomalies vanish, the residual family anomaly might still not. In fact, this is the case \emph{even if the symmetry is completely broken}. This is because we have broken the symmetry in a particular way, and we will be able to use how the symmetry relates theories at different parameter values to observe the residual family anomaly.

One situation where the residual $G_v$ anomaly determines the residual family anomaly is when $G$ acts transitively on $S(\rho)$, meaning for each $v,v'$ there is a $g \in G$ such that $g \cdot v = v'$. Indeed, if the residual $G_v$ anomaly at some $v$ vanishes, then there exists a $G_v$-symmetric nondegenerate gapping of the theory at $v$. We can then apply $G$ to that trivially gapped theory to get a uniformly gapped $G$-equivariant family on $S(\rho)$. In this case one can show $v^*$ above is an isomorphism. 

In particular, if $G$ acts freely and transitively on the $S(\rho)$, then we have an equivalence:
\begin{equation}\label{eq:guage-to-pure-gravity-anomaly}
    \Omega_{G,s,\eta}^{D+1}(S(\rho)) \cong \Omega_{s}^{D+1},
\end{equation}
where the right hand side classifies the pure gravitational anomalies in $D$ dimensions with statistic $s$. There are three examples where this occurs:
\begin{enumerate}
    \item $G = \mathbb{Z}/2$ and $\rho = \sigma$ is the $1$-dimensional sign representation.
    \item $G = U(1)$ and $\rho$ is the (real) $2$-dimensional charge $1$ representation.
    \item $G = SU(2)$ and $\rho$ is the (real) $4$-dimensional fundamental representation.
\end{enumerate}
In each cases, the family anomaly reduces to pure gravitational anomaly. We will repeatedly use this in \cref{sec:examples-of-SBLES}.

We note that in spontaneous symmetry breaking, the ground states are naturally labelled by elements of a single $G$ orbit, since degeneracy between distinct $G$ orbits may be lifted by $G$-symmetric perturbations. In this case, as above, the residual family anomaly is always determined by the anomaly of the unbroken symmetry group.

\subsubsection{$2+1$D Majoranas}\label{subsubsec2dMajoranaresidualanomaly}

Let us give a simple example of a theory with a residual family anomaly, which is nontrivial even though the symmetry is completely broken. We take a single Majorana fermion (2 component real) $\psi$ in 2+1D transforming under time reversal with $T^2 = (-1)^F$. This is known to be anomalous, and is associated with the generator of a $\bZ/16$ group of 3+1D SPTs $\Omega^4_{\rm Pin+} = \bZ/16$ \cite{Witten_2016}. This and related symmetry breaking patterns are discussed later in  \cref{subsubsecZ2fermions}.

The mass term $m \bar \psi \psi$ is $T$-odd and completely gaps the theory, so for $\sigma$ the sign representation of $\bZ/2$, the theory is $\sigma$-gappable. However, if we take $\rho = \sigma \oplus \sigma$, or equivalently the $\pi$ rotation representation of $\bZ/2$, this theory turns out not to be $\rho$-gappable. This means that for \emph{any} pair of $T$-odd operators $\cO_1$, $\cO_2$, and for any $r$, there exists a $\theta$ such that with the symmetry breaking field
\[r \cos \theta \cO_1 + r \sin \theta \cO_2,\]
the theory is not nondegenerately gapped.

\begin{figure}
    \centering
    \begin{tikzpicture}[scale=.8]
    \node at (-0.5,0) {\includegraphics[width=4cm]{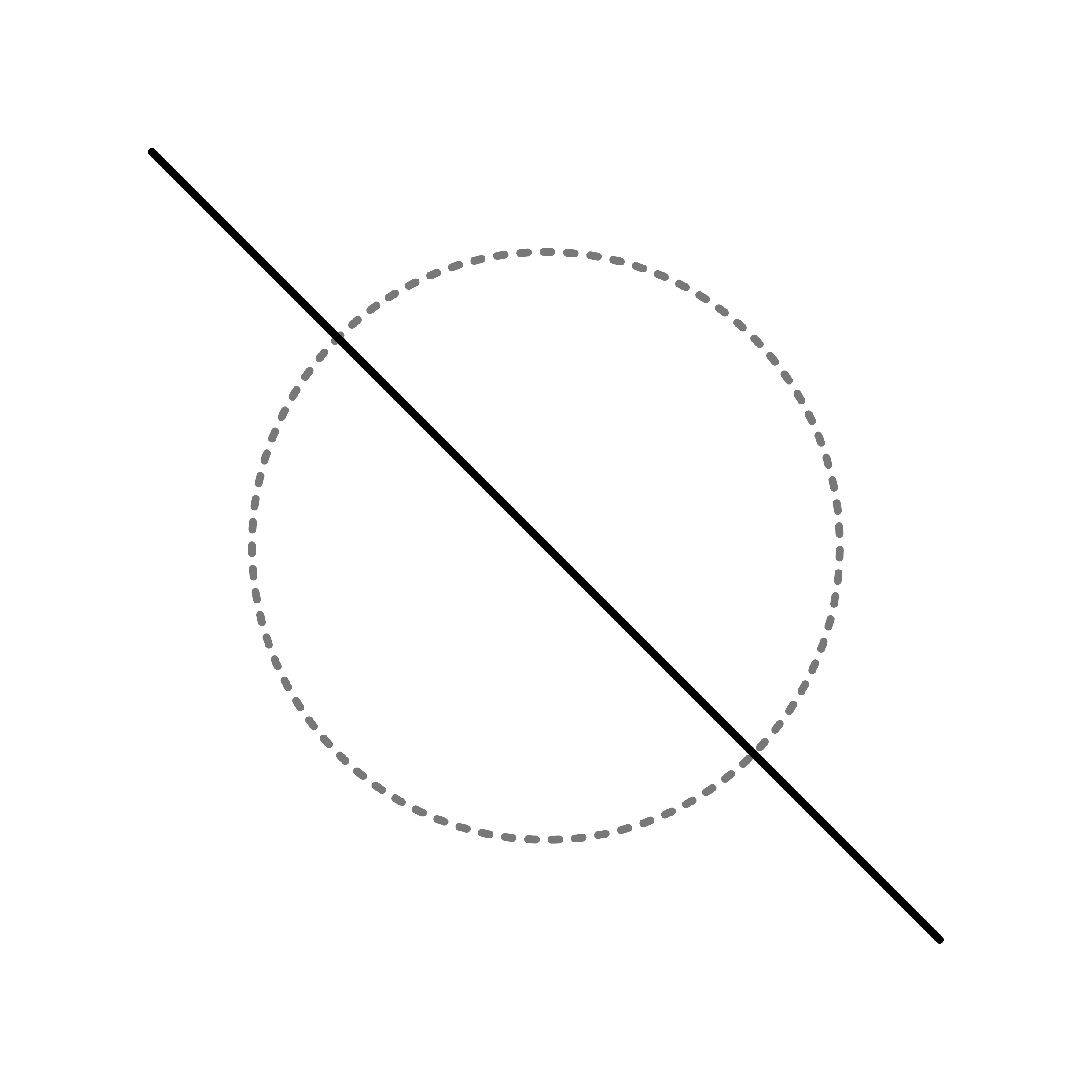}};
    \node at (4.5,0) {\includegraphics[width=4cm]{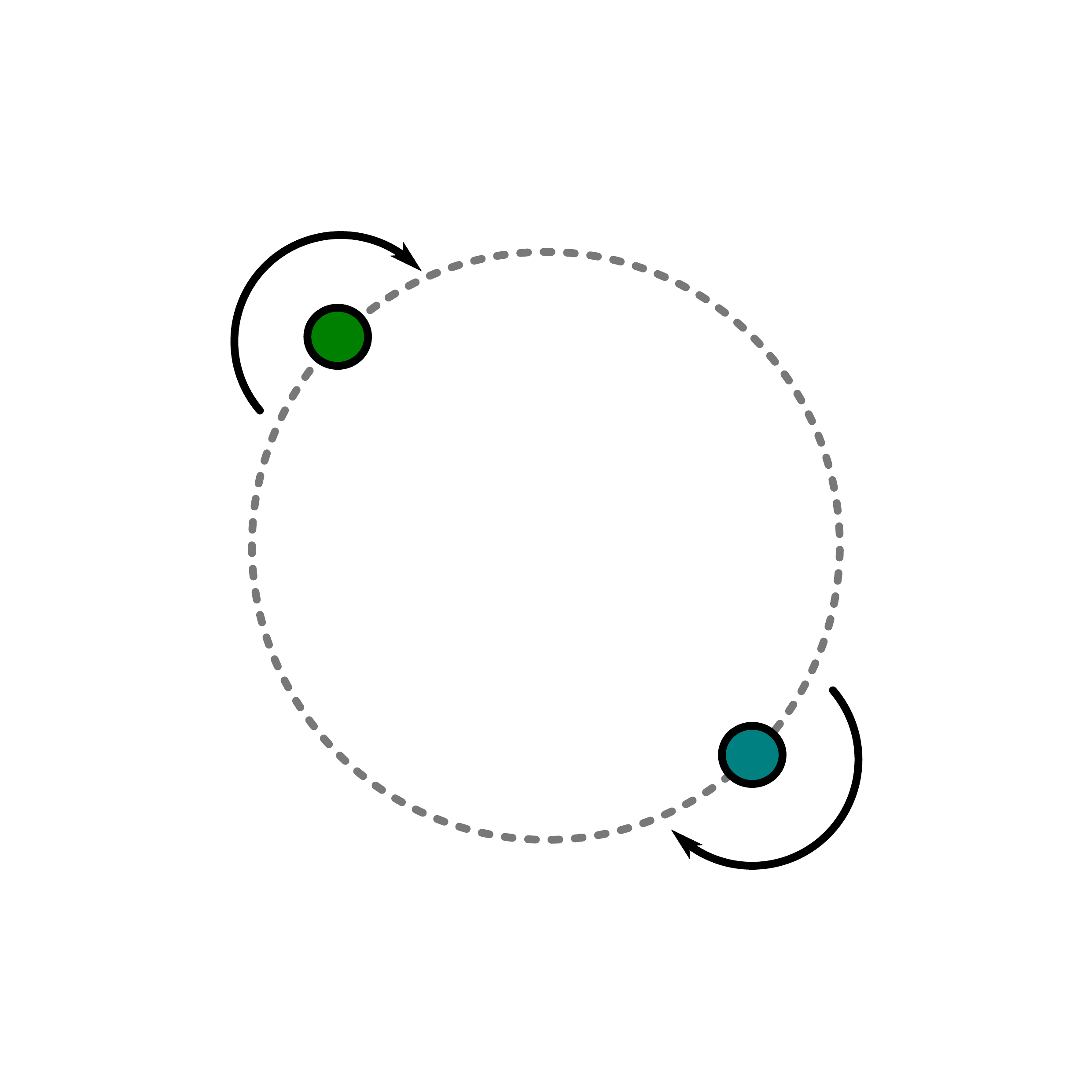}};
    \node at (-3,2) {(a)};
    \node at (2.5,2) {(b)};
    \node at (1,1.5) {$m > 0$};
    \node at (-2,-1.5) {$m < 0$};
    \node at (4.5,2) {Pump $p+ip$};
    \node at (5.5,-2) {Pump $p-ip$};
    \end{tikzpicture}
    \caption{(a) The phase diagram of a single 2+1D Majorana with redundant mass term. Time reversal acts on this phase diagram by a $\pi$ rotation. The solid black line is where the Majorana is massless, and the dotted circle represents $S(\rho)$. (b) A representation of the 3+1D invertible family, where upon crossing either the green or blue dot, a $p+ip$ or $p-ip$ superconductor is pumped to the boundary. Observe that there is no total pump in going around the entire circle. However, with time reversal, this family is non-trivial, as can be measured by going half-way around the circle and then applying time reversal to return to the starting point. The number of $p+ip$'s pumped mod 2 this way is an invariant of the equivariant family.}
    \label{fig:majoranadoublemass}
\end{figure}
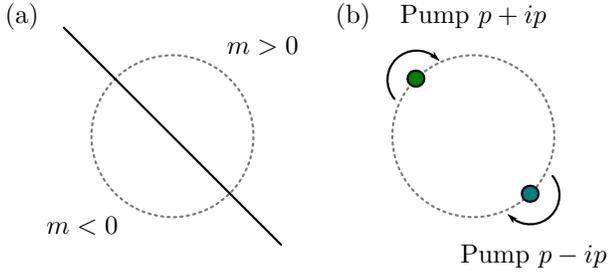

As a somewhat trivial example, if we take $\cO_1$ and $\cO_2$ to both be the (same) $T$-odd mass term, then we can always balance the coefficients so they cancel and we have the massless Majorana. This gives a phase diagram as in Fig. \ref{fig:majoranadoublemass}.

Although this phase diagram is pretty trivial, it allows us to compute the residual family anomaly. Indeed, we can observe that going around the circle by an angle of $\pi$ is equivalent to changing the sign of the mass. Majoranas with opposite mass differ by an invertible phase known as the $p+ip$ superconductor. We can say that the invertible family pumps a $p+ip$ superconductor or its inverse, a $p-ip$ superconductor, to the boundary as it crosses the $m = 0$ values of the angle; see Fig. \ref{fig:majoranadoublemass}. Observe that nothing is pumped going around the entire circle\footnote{We will see this is a general feature of families occurring in the image of the gapping obstruction in \cref{subseccompletingthecircle}.}, since the $p+ip$ and $p-ip$ are inverse phases and cancel. However, this family is still nontrivial, which can be seen as follows.

First, one can try to modify the $S^1 = S(\rho)$ family along a short arc by pumping a $p+ip$ and then a $p-ip$ at the beginning and end of said arc. However, such arcs must occur in time reversal symmetric pairs, and by inspection one can show that the number of $p+ip$'s pumped while going around half the circle is an invariant mod 2.

More precisely, in such a family we can go adiabatically half way around the circle, and then return to where we started by applying time reversal, which acts as a $\pi$ rotation. The invertible phase pumped to the boundary over such a cycle is a sort of equivariant generalization of the Thouless charge pump.

The fact that this family is nontrivial implies that the Majorana is not $\rho$-gappable for \emph{any} pair of $T$-odd operators, not just the redundant mass terms. For example we may take $\cO_1$ to be the mass term and $\cO_2$ to be any other $T$-odd operator, such as $(\bar \psi \psi)^3$.

\subsubsection{Adjoint QCD}\label{subsubsecadjointqcdexample}

Let us give a slightly more nontrivial example of a theory with a residual family anomaly, which has some interesting dynamical consequences. We consider $SU(2)$ Yang-Mills theory in 3+1D with Dirac fermions transforming in the complexified adjoint representation (equivalently we have two Majorana fermions transforming in the real adjoint). A recent discussion of this model can be found in \cite{CD18}.

There is an ABJ anomaly between the $U(1)_a$ axial symmetry and the $SU(2)$ gauge symmetry, which we can represent by the 6D integer cobordism invariant associated with (see \cref{subsecanomalyintro})
\[8 c_1^a c_2^{SU(2)} \in H^6(BU(1)_a \times BSU(2),\bZ).\]
This means that the classical $U(1)_a$ is broken down to $\bZ/8_a$ by $SU(2)$ instantons carrying 8 units of axial charge. Note that for fundamental Diracs the anomaly is $2 c_1^a c_2^{SU(2)}$ in this normalization. The relative factor of 4 can be seen by restricting to the maximal torus $U(1) < SU(2)$ for which our complex adjoint Dirac becomes a charge 2, a charge 0, and a charge $-2$ Dirac, and $2^2 + 0^2 + (-2)^2 = 8$, while for a fundamental we have $1^2 + (-1)^2 = 2$.

This theory has a 1-form $\bZ/2$ center symmetry, since the matter fields transform in the adjoint representation. If we gauge this center symmetry, it is equivalent to changing the global structure of the gauge group from $SU(2)$ to $SO(3)$. This allows for $\frac{1}{4}$ instantons that further break $\bZ/8_a$ to the $\bZ/2^F$ fermion parity subgroup (this is the same factor of 4 as above). This means there is an 't Hooft anomaly we can represent via
\[\omega = \frac{1}{4} A \Pi(B) \in H^5(B\bZ/8_a \times B^2 \bZ/2,U(1)),\]
where $A$ is the $\bZ/8_a$ gauge field, $B$ is the $B\bZ/2$ gauge field, and $\Pi(B) \in H^4(B^2\bZ/2,\bZ/4)$ is the Pontrjagin square. We will see this anomaly has a residual family anomaly for $\bZ/8_a$.

We can consider $\bZ/8_a$ chiral symmetry breaking in this theory. A natural order parameter is the charge 2 doublet consisting of the real and chiral mass terms $\bar \Psi \Psi$ and $i \bar \Psi \gamma^5 \Psi$, respectively---these form a basis of $V_\rho$. Let $\phi$ be the phase of this order parameter, which parametrizes a $\bZ/8_a$-equivariant family on $S^1$ with $\bZ/8_a$ acting as a $\pi/2$ rotation (so the $\bZ/2^F$ subgroup acts trivially).

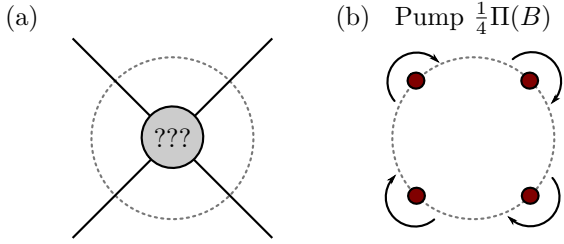
\begin{figure}
    \centering
    \begin{tikzpicture}[scale=.8]
    \node at (-0.5,0) {\includegraphics[width=4cm]{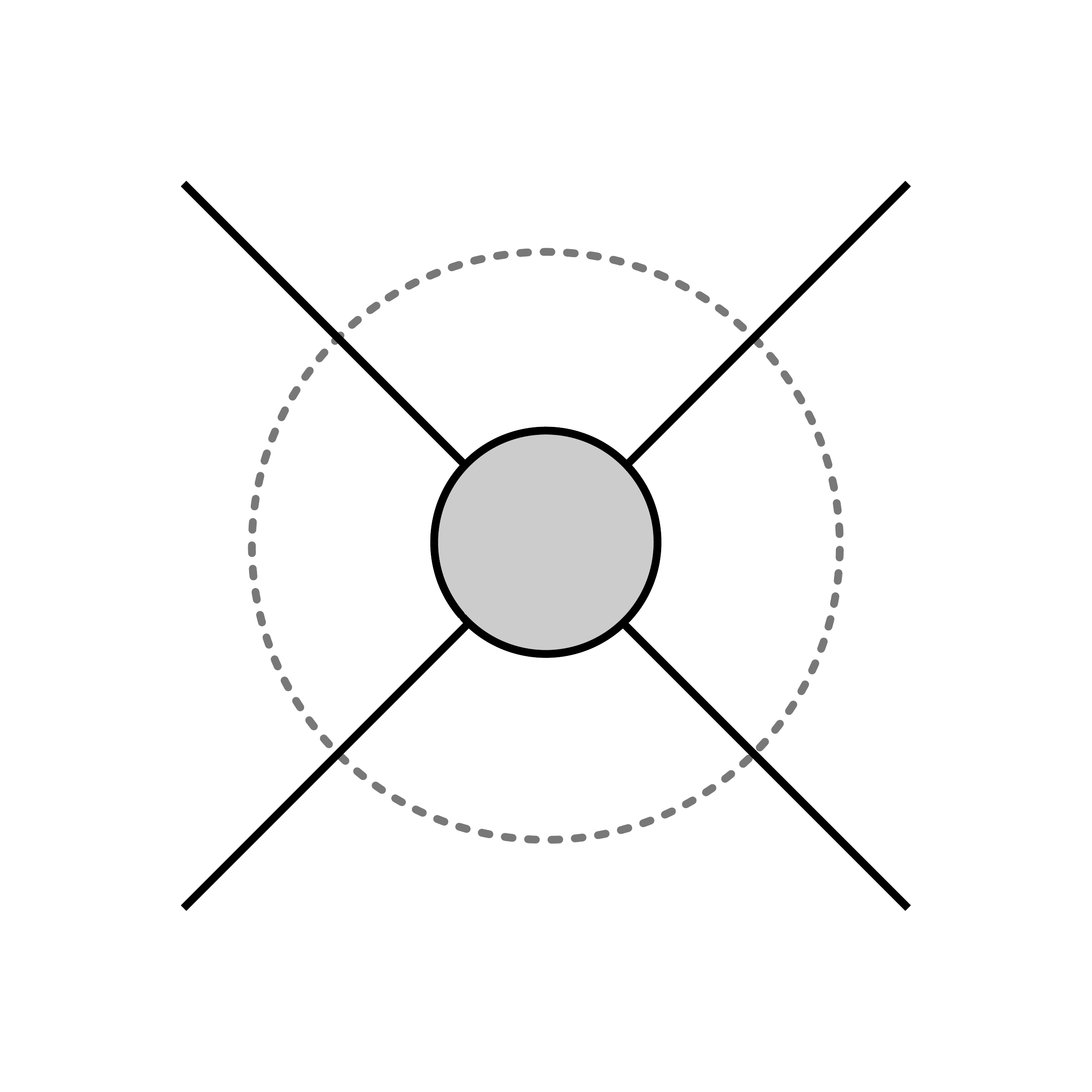}};
    \node at (4.5,0) {\includegraphics[width=4cm]{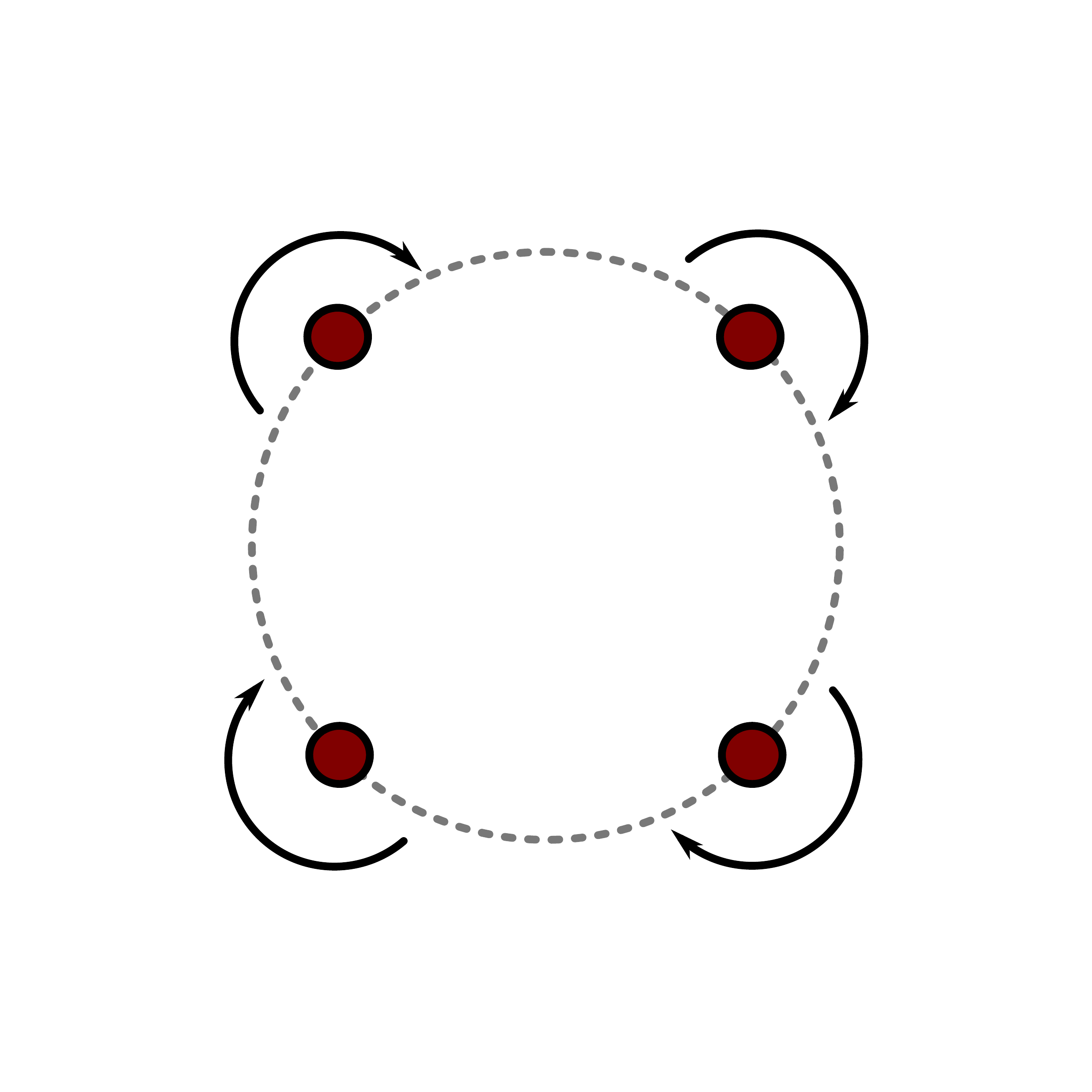}};
    \node at (-3,2) {(a)};
    \node at (2.5,2) {(b)};
    \node at (-0.5,0) {???};
    \node at (4.5,2) {Pump $\frac{1}{4} \Pi(B)$};
    \end{tikzpicture}
    \caption{(a) The phase diagram of 3+1D $SU(2)$ QCD with one adjoint Dirac fermion, deformed by the two mass terms. The $\bZ/8_a$ chiral symmetry acts as a $\pi/2$ rotation on this phase diagram. Along the four spokes, we have oblique confinement with a 2-fold degenerate ground state. At the origin, where the fermion is massless, these spokes must merge into a nontrivial point or phase. One consistent proposal is that at the origin we have $SU(2)$ chiral symmetry breaking. The deformation of this state by small masses is analyzed in Section 3.4 of \cite{CD18}. (b) The associated 4+1D invertible family over $S(\rho) \cong S^1$, where upon crossing one of the angles corresponding to oblique confinement, the 3+1D 1-form SPT $\frac{1}{4} \Pi(B)$ is pumped to the boundary. Note as in Fig. \ref{fig:majoranadoublemass} there is no total pump around the family, but if we go around by an angle $\pi/2$ and then apply the axial symmetry to return to where we started, we get a well-defined pump invariant.}
    \label{fig:adjointqcdtheta}
\end{figure}

This family is not uniformly gapped over this $S^1$. We can parametrize it by $\theta/4$, where $\theta$ is the $2\pi$-periodic QCD vacuum angle (the factor of 4 is once again the same one). However, for $\theta = \pi$, which corresponds to four different points on this $S^1$, it is expected that the theory has two degenerate ground states \cite{GKKS17,hooft1981topology}. See Fig. \ref{fig:adjointqcdtheta}.

Indeed, we can pass to the class describing the $\bZ/8_a$-equivariant $S^1$-family by replacing $A$ with $A - 4 d\phi/2\pi$, where $\phi$ is $2\pi$ periodic and parametrizes the $S^1$, since a gauge transformation by 1 shifts $A$ by 1 and $d\phi$ by $\pi/2$:
\[{\rm Res}_\rho(\omega) = \left(\frac{1}{4} A -\frac{d\phi}{2\pi}\right) \Pi(B) \in H^5_{\bZ/8_a}(S^1 \times B^2 \bZ/2,U(1)).\]
Since $\bZ/8_a$ acts freely on $S^1$ through its $\bZ/4$ quotient, we can replace $S^1$ by its quotient, parametrized by the vacuum angle $\theta = 4\phi$, and find
\[{\rm Res}_\rho(\omega) = \frac{1}{4} \frac{d\theta}{2\pi} \Pi(B) \in H^5(S^1 \times B\bZ/2^F \times B^2 \bZ/2,U(1)).\]
This is a non-trivial order 2 class, and was identified in \cite{Cordovaberry} as the family anomaly of pure QCD. See also \cite{KSTX17,GKKS17}.

\subsection{The defect anomaly matching condition}\label{subsecdefectanomap}

Let us assume now there is no residual family anomaly, i.e. ${\rm Res}_\rho(\alpha) = 0$ in \eqref{eqnssbanomaly}, and our system is nondegenerately gapped for all values of the symmetry breaking parameter $v \in S(\rho)$. In this case, we can construct a localized \textbf{$\rho$-defect} as follows. In coordinates, we take the symmetry breaking parameter $v = \phi(x)$ to vary in space, with the form
\[\phi \sim (v_1 x_1 + \cdots + v_k x_k)/\sqrt{x_1^2 + \cdots + x_k^2}\]
for large $x_1^2 + \cdots + x_k^2$,
where $v_1,\ldots,v_k$ is an orthonormal basis of $V_\rho$, so that $\phi$ winds once around $S(\rho)$ far away from a defect along $x_1 = \cdots = x_k = 0$, where it must vanish. It is crucial that the system is uniformly gapped on $S(\rho)$ for this defect to define a local $D-k$-dimensional theory.

Note that in spontaneous symmetry breaking, the topologically protected defects of codimension $n$ are classified by $\pi_{n-1}(G/H)$ \cite{sethna2021statistical}, where $H$ is the unbroken symmetry group. This coincides with the $\rho$-defect (with $k = n$) precisely in the case where $G$ acts transitively on the sphere $S(\rho)$, in which case $S(\rho) \cong G/H$ and the $\rho$ defect is a $G$-symmetric generator of $\pi_{k-1}(S(\rho)) = \mathbb{Z}$. In cases where $S(\rho)$ consists of many $G$ orbits, the $\rho$-defect is not among these topologically protected defects, but is still of interest for anomaly-matching, as we will see.

Following \cite{HKT19}, it is possible to reconstruct the 't Hooft anomaly $\alpha$ by studying the theory on the $\rho$-defect. Although the symmetry is broken, by combining the $G$ action with Lorentz symmetries (and CPT), we can invent a new symmetry $G_\rho$ (isomorphic to $G$) which acts on this effective $D-k$-dimensional theory. If the original symmetry was anomalous, there will be localized modes on the $\rho$-defect which transform nontrivially under $G_\rho$, and in particular they will have a nontrivial anomaly. We know this is the case because we can actually use this anomaly to reconstruct the bulk anomaly, as follows.

Again we use anomaly in-flow. The key is to realize that the $\rho$-defect in the anomalous $D$-dimensional theory can be extended to a $\rho$-defect in the $D+1$-dimensional $G$-SPT, so that the core of the $\rho$-defect in $D+1$-dimensions carries a $G_\rho$-SPT which controls the anomaly of the $\rho$-defect in $D$ dimensions. Thus we only need to understand how the $D+1$-dimensional SPT reduces to the $\rho$-defect.

To this end, suppose we want to compute the partition function of the $G$-SPT associated with the bulk anomaly on some $D+1$-spacetime $X$. In the presence of a symmetry breaking field $\phi$ on $X$, the $G$-SPT can be trivialized away from the $Y \subset X$ where $\phi = 0$. For generic smooth $\phi$, in the normal bundle of $Y$ we see that this zero set is precisely the bulk $\rho$-defect. Since the theory is trivialized away from $Y$, the partition function on $X$ is simply equal to the partition function of the defect anomaly theory on $Y$.

The map from spacetimes $X$ to zero sets $Y$ can be formalized, once we keep track of all the relevant structures, to define a linear map we call the \textbf{defect anomaly map}\footnote{This was called the Smith map in \cite{HKT19}, but we prefer this more descriptive name in this section.}
\[{\rm Def}_\rho\colon\Omega_{G,s,\eta+\rho}^{D+1-k} \xrightarrow{} \Omega_{G,s,\eta}^{D+1},\]
defined by
\[{\rm Def}_\rho(\alpha)(X,A)=\alpha(Y,A),\]
where $Y$ is a zero set as above. Cobordism invariance implies that this map does not depend on the choice of $\phi$ or $Y$. This map encodes the defect anomaly matching, such that if
\[\alpha \in \Omega_{G,s,\eta+\rho}^{D+1-k}\]
describes the $G$ anomaly of the $\rho$-defect, and $\omega$ our original anomaly, then we have
\[{\rm Def}_\rho(\alpha) = \omega.\]
Note that the defect anomaly $\alpha$ determines the bulk anomaly $\omega$ by this equation, but not vice versa, and in particular even anomaly-free symmetries can have anomalous $\rho$-defects, a phenomenon we will explore in \cref{subsecindexmap}.

The defect anomaly map ${\rm Def}_\rho$ and the residual family anomaly ${\rm Res}_\rho$ defined in \cref{subsecfamilyanomaly} fit together in a special way. The kernel of ${\rm Res}_\rho$ is the image of ${\rm Def}_\rho$. This means those anomalies which do not have a residual family anomaly are precisely those which can be reconstructed from the $\rho$-defect. This gives strong evidence that ${\rm Res}_\rho$ is the only obstruction to $\rho$-gappability, since we used this to define the $\rho$-defect. It also generalizes Theorem 4.2 in \cite{HKT19} from finite cyclic groups to arbitrary groups and arbitrary representations, answering the question of the cokernel of ${\rm Def}_\rho$ (i.e.\ the Smith map) which was posed there.

\subsubsection{3+1D Dirac fermion}\label{defect_match_Dirac}

Consider a 3+1D Dirac fermion $\psi$ (with four complex components). This has an anomalous chiral symmetry $U(1)_L$ which gives charge 1 to the two left-handed components of $\psi$ and charge 0 to the two right handed ones. There are two Dirac masses $\bar \psi \psi$ and $i \bar \psi \gamma^5 \psi$, which transform together under $U(1)_L$ as a charge 1 doublet $\rho$. Any combination of the two mass terms completely gaps the fermion, so in this case there is no residual family anomaly and there is a local $\rho$-defect.

We can construct the $\rho$-defect in this theory by choosing a spatially-varying mass profile of the form
\[x_1 \bar \psi \psi + x_2 i\bar \psi \gamma^5 \psi.\]
One can solve the Dirac equation for localized modes with this mass profile and find a massless 1+1D Weyl fermion (with one complex component) propagating in the remaining coordinates \cite{callan1985anomalies}. We will evaluate ${\rm Def}_\rho$ for 1+1D theories with this symmetry and show by anomaly-matching that this fermion must have charge 1 under the residual $U(1)_\rho$ symmetry (which could also be concluded by a careful analysis of the localized solutions).

The residual symmetry $U(1)_\rho$ acting on the 1+1D $\rho$-defect acts as a combination of a $U(1)_L$ rotation and a compensating $\Spin(2)$ rotation, where $\Spin(2)$ is the rotation in the $x_1,x_2$ plane, such that the mass profile is invariant under their combination. In particular, a $2\pi$ $U(1)_\rho$ rotation is equal to a $2\pi$ rotation of this plane, which equals the fermion parity $(-1)^F$. This means we are interested in 1+1D systems with $\Spin^c = (\Spin \times U(1)_\rho)/\bZ/2$ structure. A general anomaly for such a theory is given by a Chern-Simons form associated with a 4D integer cobordism invariant (see \cref{subsecanomalyintro})
\[\label{eqn2dspincanom}\alpha = k_1 \left(\frac{1}{8}(c_1^\rho)^2 - \frac{1}{24} p_1(TY)\right) + k_2 (c_1^\rho)^2,\]
where $k_1, k_2 \in \bZ$.

We can compute ${\rm Def}_\rho(\alpha)$ in terms of these 4D cobordism invariants. That is, suppose $X$ is a closed 6D Spin manifold with a principal $U(1)_L$ bundle $P$ and a section $\phi$ of the $\mathbb{C}$ bundle $E_\rho \coloneqq P \times_{U(1)_L} V_\rho$ associated to the charge 1 representation $V_\rho$. We take $Y$ to be the analog of the $\rho$-defect, i.e. it is the zero set of $\phi$ (we can always perturb $\phi$ so its zero set is a 4-manifold). A useful fact is that the homology class $[Y] \in H_4(X,\bZ)$ is Poincar\'e dual to the first Chern class $c_1^L \in H^2(X,\bZ)$. This means that for any $\beta \in H^4(X,\bZ)$,
\[\int_X c_1^L \beta = \int_Y \beta.\]
To compute ${\rm Def}_\rho(\alpha)$, we want to choose $\beta$ such that $\beta|_Y = \alpha$, then by definition we will have ${\rm Def}_\rho(\alpha) = c_1^L \beta$.

To get the $(c_1^\rho)^2$ terms, we use the fact that the $U(1)_\rho$ bundle over $Y$ is defined by restriction of the $U(1)_L$ bundle, so in particular $c_1^L|_Y = c_1^\rho$. In terms of the defect anomaly map, this means to get $(c_1^\rho)^2$ we should take $\beta = (c_1^L)^2$, so
\[\label{eqndefanompureint2dmap}{\rm Def}_\rho((c_1^\rho)^2) = (c_1^L)^3.\]

The ``gravitational'' term (involving $p_1(TY)$) is more interesting. If we study the tangent bundle of $X$ restricted to $Y$ we find
\[TX|_{Y} = TY \oplus NY = TY \oplus E_\rho|_Y,\]
where we have identified the normal bundle $NY$ with the restriction of the associated bundle $E_\rho$, since $Y$ is the zero set of the section $\phi$. Using the Whitney sum formula we obtain
\begin{equation}
    \begin{aligned}
        p_1(TX)|_Y &= p_1(TY) + p_1(E_\rho)|_Y \\ &= p_1(TY) + (c_1^L)^2|_Y\\ 
        &= p_1(TY) + (c_1^\rho)^2.
    \end{aligned}
\end{equation}
So to get
\[\alpha = \frac{1}{8} (c_1^\rho)^2 - \frac{1}{24} p_1(TY)\]
we should take
\[\beta = \frac{1}{6} (c_1^L)^2 - \frac{1}{24} p_1(TX),\]
hence
\[{\rm Def}_\rho\left(\frac{1}{8}(c_1^\rho)^2 - \frac{1}{24} p_1(TY)\right) = \frac{1}{6} (c_1^L)^3 - \frac{1}{24} c_1^L p_1(TX).\]

This turns out to precisely coincide with the $U(1)_L$ anomaly of the 3+1D Dirac fermion. Thus defect anomaly matching requires $k_1 = 1$, $k_2 = 0$ in \eqref{eqn2dspincanom}. This is consistent with a 1+1D Weyl fermion with $U(1)_\rho$ charge 1.

The above calculation seems to rely on a choice of $\beta$. Actually, it does not, since if $\beta'|_Y = \alpha$, $(\beta - \beta')|_Y = 0$, and so, using Poincar\'e duality, $c_1^L \beta - c_1^L \beta' = c_1^L (\beta - \beta') = 0$. On the other hand, the existence of such a $\beta$ is guaranteed by the vanishing of the residual family anomaly, since this guarantees that $\int_X \omega = \int_Y \beta$ for some $\beta$.

\subsubsection{3+1D Weyl fermion}\label{3+1dWeyl}

To see the importance of the representation in the above computation, let us consider a closely related example, this time beginning with a left-handed Weyl fermion in $3+1$D. This has a $U(1)_L$ symmetry with the same anomaly as the Dirac in \cref{defect_match_Dirac} (since the right-handed Weyl does not contribute anything):
\begin{equation}
    \omega  = \frac{1}{6}(c_1^L)^3 - \frac{1}{24}c_1^Lp_1(TX).
\end{equation}

Above we studied the Dirac mass, which couples the two Weyl components. However, a single Weyl on its own has a Majorana mass that is \emph{charge 2} under $U(1)_L$. 
Solving the equations of motion for the associated $\rho$-defect we find a left-handed Majorana-Weyl fermion in 1+1D. This has one real component, so $U(1)_\rho$ must act trivially on it.

Let us compute the defect anomaly map in this case and verify that this matches.
Note that a $2\pi$ rotation in $U(1)_\rho$ is a $4\pi$ rotation in $V_\rho$, which is 1 on the fermion, so there is no $\Spin^c$ business here. Anomalies of 1+1D fermions with $\Spin \times U(1)_\rho$ symmetry split between a pure gravity and a pure symmetry part, and take the form
\[\alpha = \frac{k_1}{48} p_1(TY) + k_2 (c_1^\rho)^2.\]
The calculation proceeds as above, although now $[Y] \in H_4(X,\bZ)$ is Poincar\'e dual to $2 c_1^L \in H^2(X,\bZ)$, since $\rho$ is a charge 2 representation. Once we compute $\beta$ such that $\beta|_Y = \alpha$, we will have ${\rm Def}_\rho(\alpha) = 2 c_1^L \beta$.

Using $c_1^L|_Y = c_1^\rho$, we find
\[{\rm Def}_\rho((c_1^\rho)^2) = 2 (c_1^L)^3.\]
We also have
\begin{equation}
    \begin{aligned}
      p_1(TX)|_Y &= p_1(TY) + p_1(E_\rho)|_Y\\
                 &= p_1(TY) + 4 (c_1^L)^2|_Y \\
                 &= p_1(TY) + 4 (c_1^\rho)^2.  
    \end{aligned}
\end{equation}
Thus we find
\[{\rm Def}_\rho\left(\frac{1}{48} p_1(TY)\right) = \frac{1}{6} (c_1^L)^3 - \frac{1}{24} c_1^L p_1(TX),\]
so the defect anomaly matches correctly with $k_1 = 1$, $k_2 = 0$.

\subsection{The index map and higher Berry phase}\label{subsecindexmap}

Above we described an anomaly matching condition in terms of a map ${\rm Def}_\rho$ for which the image of the defect anomaly $\alpha$ is the bulk anomaly $\omega$:
\[{\rm Def}_\rho(\alpha) = \omega.\]
We see the defect anomaly determines the bulk anomaly, but when ${\rm Def}_\rho$ is not injective, there can be several solutions for $\alpha$ given $\omega$. Thus there is an ambiguity in the defect anomaly. There can even be anomalous defects ($\alpha \neq 0$) in anomalous bulk theories ($\omega = 0$)!

Recall that as long as there is no residual family anomaly, we can perturb things so that for each large enough value of the symmetry-breaking field, we obtain a trivially gapped ground state. This defines a $G$-equivariant family of invertible field theories over the sphere $S(\rho)$. This family is not typically free of $G$-anomalies, but it is when $\omega = 0$. In this case, we can couple it to a $G$ gauge field, and classify its topological response by an element
\[\zeta \in \Omega^D_{G,s,\eta}(S(\rho))\]
(cf. \cref{subsecanomalyintro}). Given such a family, we can construct the $\rho$-defect as before, and we want to describe the anomaly.

\begin{figure}
    \begin{tikzpicture}
    \node at (0,0) {\includegraphics[width=7cm]{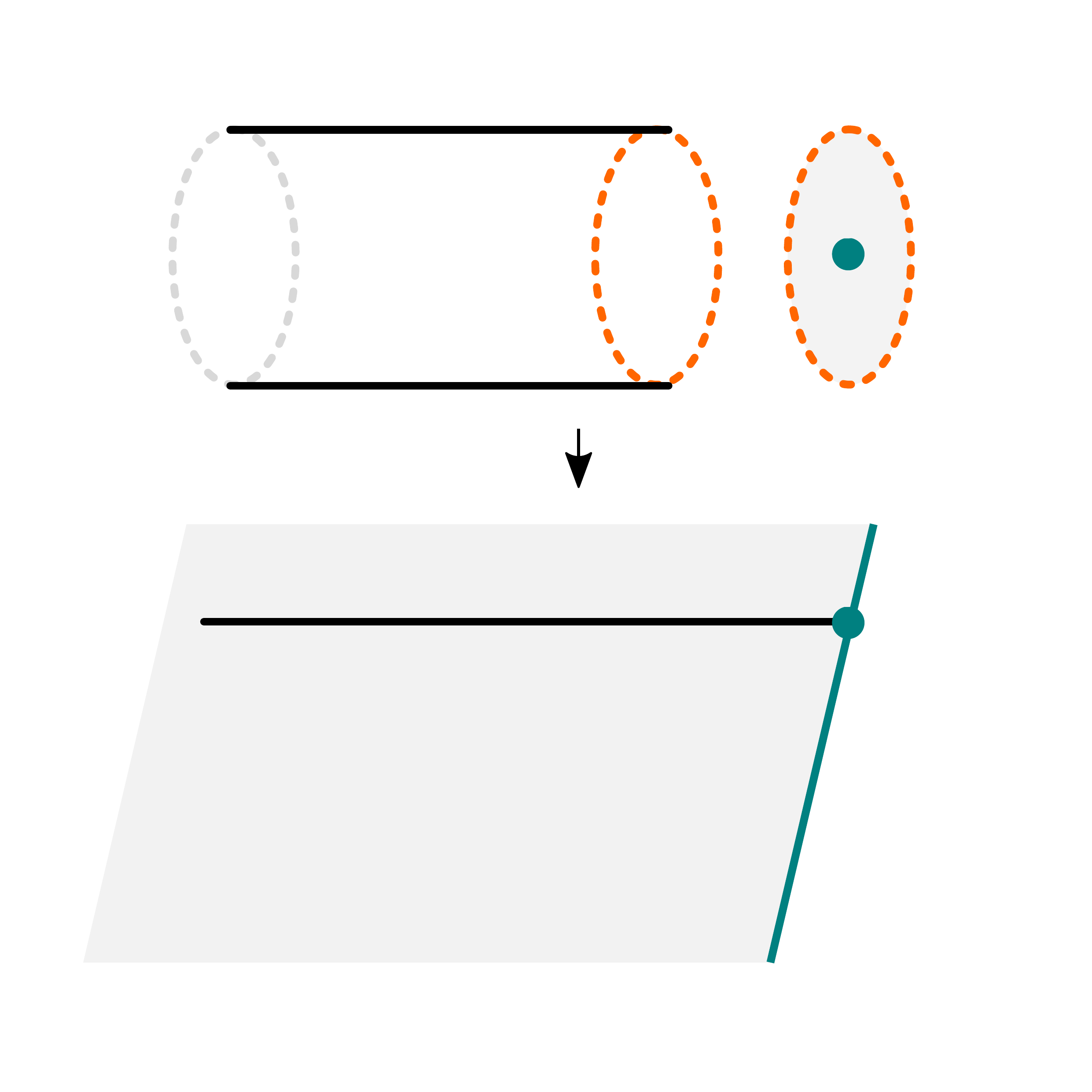}};
    \node at (2.7,2) {$B^k$};
    \node at (0.6,3.2) {$S^{k-1}$};
    \node at (2.8,-1.3) {$\rho$-defect};
    \node at (-1,-1.3) {${\rm Ind}_\rho$ theory};
    \end{tikzpicture}
    \caption{\textbf{Calculating the index map:} the index map ${\rm Ind}_\rho$ describes the anomaly of a $\rho$-defect inside an invertible phase via a certain sphere compactification of that phase described in the text. The proof-by-picture of why this works is given here. The $\rho$-defect is defined on $B^k \times \bR^{D-k}$, where $B^k$ is a $k$-dimensional ball, depicted here as a $B^k$ bundle over $\bR^{D-k}$ (blue). Meanwhile we consider the invertible phase defined on $S^{k-1} \times H^{D-k+1}$, where $H^{D-k+1}$ is a $D-k+1$-dimensional half-space, shown as an $S^{k-1}$ bundle over $H^{D-k+1}$ (gray). These have the same boundary (orange), and can be glued together to define a boundary condition of the compactified invertible theory, so long as the order parameter winds around this $S^{k-1}$. This defines ${\rm Ind}_\rho$ and thus measures the anomaly of the $\rho$-defect by anomaly in-flow.}
    \label{fig:index-map}
\end{figure}

We can actually construct the anomaly theory of the $\rho$-defect directly from $\zeta$ by compactifying on $S(\rho) \cong S^{k-1}$. The idea is shown in Fig. \ref{fig:index-map}. The compactification defines an element of $\Omega_{G,s,\eta+\rho}^{D+1-k}$, and moreover we get the \textbf{index map}
\[{\rm Ind}_\rho \colon \Omega^D_{G,s,\eta}(S(\rho))  \to \Omega_{G,s,\eta+\rho}^{D-k+1}.\]
In terms of partition functions, this map is defined as follows. Suppose we have a $(D-k+1)$-dimensional spacetime $Y$, equipped with a $G$-connection $A$ and $(\eta+\rho)$-twisted $s$-structure $\xi$. We can define the $D$-dimensional spacetime $W$ given as the total space of the $S(\rho)$-bundle over $W$ associated to the $G$ gauge bundle.

$W$ gets a $G$ connection $\pi^*A$ by pullback from the projection map $\pi \colon W \to Y$. Its tangent bundle can be identified with $\pi^*TY \oplus \pi^*A^*V_\rho = \pi^*(TY \oplus A^*V_\rho)$, so it gets an $\eta$-twisted $s$-structure $\pi^*\xi$ from the $\eta+\rho$-twisted $s$-structure on $Y$. Finally, if we study the $S(\rho)$-bundle on $W$ associated to $\pi^* A$, we can write the total space of this bundle as
\[\pi_W:(S(\rho) \times S(\rho)) \times_G P_A \to W = S(\rho) \times_G P_A,\]
where $P_A$ is the gauge bundle of $A$. The map $\pi_W$ is given by projecting to the first $S(\rho)$ factor, and we get a section $\phi_W$ by the diagonal map $S(\rho) \to S(\rho) \times S(\rho)$. Thus, given an element $\zeta \in \Omega^D_{G,s,\eta}(S(\rho))$, we can define
\[\label{eqnindexmapdef}{\rm Ind}_\rho(\zeta)(Y,A,\xi) = \zeta(W, \pi^* A, \pi^*\xi,\phi_W).\]

We can also consider elements of $\Omega^D_{G,s,\eta}(S(\rho))$ as $D$-dimensional counterterms which can appear relating different symmetry-breaking patterns of a given theory with the same representation $\rho$. In particular, we can compare two different $G$-equivariant $S(\rho)$-families of invertible field theories by stacking one with the orientation reversal of the other. The result is free of $G$-anomalies and defines an element of $\Omega^D_{G,s,\eta}(S(\rho))$. Thus, the image of ${\rm Ind}_\rho$ above describes both the ambiguity in the defect anomaly and the kernel of ${\rm Def}_\rho$ (answering the question of the kernel of the Smith homomorphism in \cite{HKT19}).

The index map can be thought of as a generalization of the Callias index theorem \cite{callias1978axial,bott1978some} which computes the fermion zero modes at the core of a mass defect. Our map gives the $G$-anomaly of those zero modes (and thus accounts for interactions).

If we define $B(\rho)$ as the ball in $V_\rho$ with boundary $S(\rho)$, the index map is the obstruction to extending the $S(\rho)$ family to a $G$-equivariant family on $B(\rho)$.
In particular, the point $0 \in B(\rho)$ is a $G$-symmetric invertible field theory, and therefore the kernel of ${\rm Ind}_\rho$ is the image of ${\rm Res}_\rho$! In terms of bulk-boundary correspondence, the index map is the obstruction to a $G$-equivariant family admitting a $G$-symmetry boundary condition which is \emph{independent} of the parameters.

\subsubsection{Thouless pump and vortices}\label{subsubsecexamplethouless}

We will consider the relationship between the index map and the Thouless pump. We begin with a 1+1D Dirac fermion (with two complex components) with its anomaly-free $U(1)$ symmetry
\[\psi \mapsto e^{i\theta/2} \psi.\]
Suppose we add a $U(1)$-symmetric mass term
\[i((\cos \phi) \bar \psi \psi + i (\sin \phi) \bar \psi \gamma^c \psi),\]
where $\gamma^c$ is the chirality operator $i\gamma^0 \gamma^1$. This defines a $U(1)$-\emph{symmetric} $S^1$-family of invertible field theories parametrized by $\phi$. This family is nontrivial, and can be described by
\[\label{eqnzetathouless}\zeta(W,A,\phi) = \frac{1}{2\pi} \int_W d\phi A,\]
where $W$ is the 1+1D spacetime, $A$ is a Spin$^c$ structure, and $\phi \colon W \to S^1$. As described in \cref{subsecanomalyintro}, the physics of this term is we get an $A$ current when adiabatically varying the $S^1$ parameter, leading to a quantized charge pump (the classic Thouless pump \cite{thouless1983quantization}).

We expect the $\rho$-defect, which is the operator which creates a vortex in $\phi$, to carry a unit $A$ charge which matches the Thouless pump. This will be the result of the index map, which in this case takes
\[{\rm Ind}_\rho\colon \bZ \cong \Omega^2_{\Spin^c}(S^1) \to \Omega^1_{\Spin^c} \cong \bZ,\]
where the latter group can be thought of as the group of $A$ charges. Note that since the image of ${\rm Ind}_\rho$ is the kernel of ${\rm Def}_\rho$, and $\rho$ here is trivial (we have a symmetric family) so ${\rm Def}_\rho = 0$, we already know on abstract grounds that this map is surjective, and hence an isomorphism. Let us compute it to check.

To compute the map, we use \eqref{eqnindexmapdef}. That is, we will associate to $\zeta$ in \eqref{eqnzetathouless} a partition function of 0+1D spacetimes $Y$ (which are merely collections of oriented circles) equipped with a Spin$^c$ connection $A$. We start by forming the associated $S(\rho)$ bundle over $Y$. Since $\rho$ is trivial, this bundle is simply a product $W = S^1 \times Y$. The canonical section $\phi\colon W \to W \times S^1$ is the product of diagonal map $S^1 \to S^1 \times S^1$ and the identity map $Y \to Y$. In particular, $d\phi/2\pi$ is the volume form on the $S^1$ factor. It follows
\begin{equation}
    \begin{aligned}
        {\rm Ind}_{\rho}(\zeta)(Y,A,\phi) &= \zeta(W,\pi^*A,\phi) \\
        &= \frac{1}{2\pi} \int_W d\phi \,\pi^* A \\
        &= \int_{S^1} \frac{d\phi}{2\pi} \int_Y A 
       \quad  = \int_Y A,
    \end{aligned}
\end{equation}
which is the generator of $\Omega^1_{\Spin^c}$, as expected.

\subsubsection{Berry phase and projective representations}\label{subsubsecberryphaseprojrep}

We study the relationship between projective symmetry and Berry phase via the index map. 

Let us take $G = SO(3)$ acting on a Hilbert space carrying spin $s/2$, initially with $H = 0$. We can think of this as a $D=1$ system with anomaly
\[\omega = \frac{1}{2} s w_2 \in H^2(BSO(3),U(1)) \cong \bZ/2,\]
where $w_2$ is the generator of $H^2(BSO(3),U(1))$.

We then apply a ``magnetic field''
\[H(B) = - \vec{B} \cdot \vec{S}\]
to this spin. The parameter $\vec{B} \in \bR^3$ transforms in the adjoint representation $\rho$ of $SO(3)$, and so for any nonzero value, $SO(3)$ is broken down to the $SO(2)$ subgroup of rotations around the $\vec B$ axis. Furthermore, for any nonzero value, $H(B)$ has a unique ground state. This means that the residual anomaly
\[{\rm Res}_\rho (\omega) = 0,\]
and thus we expect $\omega$ to be in the image of the defect anomaly map. 

The defect anomaly lives in
\[\Omega^{-1}_{SO(3),SO} = H^0(BSO(3),\bZ) = \bZ,\]
and so evidently
\[{\rm Def}_\rho\colon  \bZ \to \bZ/2\]
is reduction mod 2. However, the interpretation of the defect anomaly is not obvious, since it seems to encode an anomaly of a $-2$-spacetime-dimensional system. The correct interpretation of this $\bZ$ (which follows from the definition of ${\rm Def}_\rho$) is the Chern number of the Berry bundle over $S(\rho) \cong S^2$ family, which is known to equal the spin $s$, consistent with the anomaly above.

The index map is
\[{\rm Ind}_\rho \colon\Omega^2_{SO(3),SO}(S(\rho)) \cong \bZ \to \bZ,\]
which by exactness must be multiplication by 2, since its image is the kernel of the quotient ${\rm Def}_\rho \colon\bZ \to \bZ/2$. We can interpret this map as follows. Suppose the spin $s/2$ is an integer, so we are in the kernel of ${\rm Def}_\rho$, meaning there is no anomaly and the Hilbert space carries an honest representation of $SO(3)$. 

We can generalize the magnetic field Hamiltonian above, which projects onto a highest weight vector, to one which projects onto a vector of weight $l$ (the magnetic quantum number). For each $l \in \{-s/2, -s/2+1, \ldots, s/2\}$, this Hamiltonian transforms in the adjoint of $SO(3)$. We find $l$ is encoded in the $SO(3)$-equivariant $S^2$ family as the charge of the unbroken $SO(2)$ at any fixed value. This family thus represents $l \in \Omega^2_{SO(3),SO} = \bZ$ via the isomorphism
\[\Omega^2_{SO(3),SO} = H^2_{SO(3)}(S^2,\bZ) = H^2(BSO(2),\bZ),\]
where the latter represents the charge of the unbroken $SO(2)$ at a fixed value (see the discussion in \cref{subsecfamilyanomaly} about transitive group actions). Indeed, it is known in this case that the Chern number of the resulting Berry connection is $2l$, which agrees with the index map above.

We note that for representations $\rho$ of dimension greater than 3, since $\Omega^{2-k}_{G,s,\eta} = 0$ for all $k > 3$ and all $G$, $s$, $\eta$, if there is a projective representation, there is no Berry phase that can match this anomaly by ${\rm Def}_\rho$. Since the image of ${\rm Def}_\rho = 0$ is the kernel of ${\rm Res}_\rho$, the residual family anomaly map is therefore \emph{injective}. In particular, the family is not uniformly gapped over $S(\rho)$.

For example, suppose we take $G = PSU(n)$, with our Hilbert space corresponding to the $SU(n)$ vector representation. This is a projective $PSU(n)$ representation and has anomaly generating the group
\[\omega = \frac{1}{n} u_2 \in H^2(BPSU(n),U(1)) \cong \bZ/n.\]
The spin-$1/2$ case above corresponds to $n = 2$, via $PSU(2) = SO(3)$. The analog of the magnetic field Hamiltonian above is
\[H(B) = -\sum_i B^i S_i\]
where $S_i \in su(n)$ ranges over a basis of the traceless Hermitian $n \times n$ matrices. As before, $\vec B$ transforms in the adjoint representation $\rho$ of $PSU(n)$, which has dimension $n^2 -1$.

When $n > 2$, the Hamiltonian $H(B)$ does not have a unique gapped ground state for all $B \neq 0$. The issue is that the lowest two (or more, up to $n - 1$) eigenvalues of $H(B)$ may be degenerate, while the other eigenvalues can balance them so ${\rm Tr\ }H(B) = 0$, without making $H(B)$ identically zero. We anticipated this based on the long exact sequence, and indeed there is a residual family anomaly, which generates the group
\[{\rm Res}_\rho (\omega) \in H^2_{PSU(n)}(S(\rho),U(1)) \cong \bZ/n.\]
To see this, we observe that if we take $B$ to be one of the points in $S(\rho)$ with two degenerate lowest energy states, there is an unbroken $PSU(2)$ with $\bZ/n$ anomaly $\frac{1}{n} u_2$, which must be given by $B^* {\rm Res}_\rho (\omega)$ (cf. \cref{eqnpullbackfamily} and Eq. \eqref{eqnpullbackfamily}).

\subsubsection{Time reversal domain wall for 2+1D Majorana fermions}\label{subsubsectdwformajorana}

Let us analyze an example from \cite{HKT19} of a situation with ambiguous defect anomaly. We study $N_f$ 2+1D Majorana fermions $\psi_j$ with time reversal
\[T\psi_j = \gamma^0 \psi_j,\]
which satisfies $T^2 = (-1)^F$. This has an anomaly $\omega = N_f \omega_4 \in \Omega^4_\Spin(B\bZ/2,3\sigma) = \Omega^4_{{\rm Pin}^+} \cong \Z/16$, where $\omega_4$ is the generator corresponding to $N_f = 1$ (it can be expressed as an eta invariant of the Dirac operator \cite{WittenParity}). This example is also a member of the 4-periodic family discussed later in \cref{subsubsecZ4fermions}.

Let us consider $N_f = 2$. Time reversal can be broken by mass terms such as
\[\bar \psi_1 \psi_1 \pm \bar \psi_2 \psi_2.\]
(Each $T$-odd mass term transforms in the sign representation, which is $\rho$ here.) On the time reversal domain wall there is a unitary $\Z/2$ symmetry $U$, whose anomaly group is classified by $\Omega^3_{\Spin,\Z/2} \cong \Z \oplus \Z/8$, the first part $\alpha_3$ being purely gravitational and the second part $\alpha_3^{\bZ/2}$ involving the internal symmetry $U$. It turns out that depending on the relative sign, the domain wall has different anomalous modes. If the sign is the same, on the wall we have two 1+1d Majorana modes of the same chirality. However, if we take opposing signs, we get two Majoranas with opposite chirality. These clearly have distinct gravitational anomalies, and it turns out they have distinct $U$ anomalies as well, with $U$ acting trivially in the first case and chirally in the second case.

Although they have different anomalies, both must satisfy the defect anomaly matching condition. Since ${\rm Def}_\rho$ is linear, we can use the two data points above to compute it, and find, in terms of generators $k_1 \in \Z$, $k_2 \in \Z/8$,
\[{\rm Def}_\rho(k_1 \alpha_3 + k_2 \alpha_3^{\bZ/2}) = (k_1 - 2k_2)\omega_4,\]
where $(k_1,k_2)$ is $(2,0)$ or $(0,1)$ in the two domain walls above, and both match the anomaly $2\omega_4$ as expected.

We see that the kernel of ${\rm Def}_\rho$ is generated by $(2,1)$, which was noted in \cite{HKT19}. We can see ambiguity arising from ${\rm Ind}_\rho$ as follows. We need to start by considering 2+1D $\bZ/2^T$-equivariant families of invertible field theories over $S(\rho)$. In this case, $S(\rho) = S^0$ is just two points which get exchanged by $T$. The generator $\zeta \in \Omega_{\bZ/2,\rm Spin,3\sigma}^3(S(\rho)) = \bZ$ is defined by taking the generator $\alpha_3 \in \Omega_\Spin^3 = \bZ$ over one of the two points, and its time-reversed partner $-\alpha_3$ over the other point.

To calculate ${\rm Ind}_\rho$, we study the interface between these two invertible theories. The result is two fermions of equal chirality (gravitational anomaly $2\alpha_3$), which are swapped by the induced $\Z/2$ symmetry $U$. This swap has eigenvalues $\pm 1$ and we find its anomaly is $\alpha_3^{\bZ/2}$. So if $\zeta$ is the class of the family above,
\[{\rm Ind}_\rho(\zeta) = 2\alpha_3 + \alpha_3^{\bZ/2},\]
the image of which is indeed the kernel of ${\rm Def}_\rho$ we computed above.

This has a physical interpretation in terms of the two mass terms above. If we change the sign of just the $\bar \psi_2 \psi_2$ mass term, we can think of this as stacking with either $\alpha_3$ or $-\alpha_3$, depending whether the sign change is from minus to plus or from plus to minus. This gives the invertible family $\zeta$ above.

\subsubsection{Vortices in $p+ip$ superfluid}\label{subsubsecpipsuperfluid}

Now we will discuss the famous Majorana zero modes bound to the vortices of a $p+ip$ superfluid \cite{volovik2003universe}, which turn out to have an interesting description in terms of the index map.

We study a single Dirac fermion in 2+1D, carrying charge 1 under $G = U(1)$ symmetry, and undergoing symmetry breaking via a charge 2 complex order parameter coupling to the two Majorana masses. Such a spontaneous symmetry breaking scenario is typically referred to as a $p+ip$ superfluid.\footnote{Note that there is a mixed $U(1)$ and time reversal anomaly, and a choice of $U(1)$ symmetric fermion regulator will break time reversal and select either a $p+ip$ or $p-ip$ superfluid.}
The resulting $S(\rho) \cong S^1$ family has a unique gapped ground state for all nonzero values of the order parameter, and the $U(1)$ symmetry is anomaly-free, and it represents a generator of
\[\Omega^3_{\Spin,U(1),\rho}(S(\rho)) \cong \bZ.\]
We want to compute the index map
\[{\rm Ind}_\rho^{U(1)}\colon\Omega^3_{\Spin,U(1),\rho}(S(\rho)) \to \Omega^2_{\Spin,U(1)} \cong \bZ/2.\]

It is interesting to consider the map
\[f\colon\Omega^3_{\Spin,U(1),\rho}(S(\rho)) \to \Omega^3_{\Spin}(S^1) \cong \bZ \oplus \bZ/2\]
which forgets the $U(1)$ action, since the index map of the latter, namely
\[{\rm Ind_\rho}\colon\Omega^3_{\Spin}(S^1) \to \Omega^2_\Spin\]
can be more easily understood. The generators of $\Omega^3_{\Spin}(S^1)$ correspond to the generator of $\Omega^3_\Spin \cong \bZ$, with trivial parameter dependence, and $\Omega^2_\Spin = \bZ/2$, via a family which pumps this phase to the boundary as we go around $S^1$. The index map clearly sends the $\bZ$ generator to zero and the $\bZ/2$ generator to the generator of $\Omega^2_\Spin = \bZ/2$.

Because the SBLES is functorial in $G$, we have a commutative square
\begin{center}
\begin{tikzcd}
\bZ \cong \Omega^3_{\Spin,U(1),\rho}(S(\rho)) \arrow[r, "{\rm Ind}_\rho^{U(1)}"] \arrow[d,"f"]
&  \Omega^2_{\Spin,U(1)} \cong \bZ/2 \arrow[d, "\sim"] \\
\bZ \oplus \bZ/2 \cong \Omega^3_{\Spin}(S^1) \arrow[r, "{\rm Ind}_\rho"]
&  \Omega^2_\Spin \cong \bZ/2.
\end{tikzcd}
\end{center}
Combined with the information above, we learn ${\rm Ind}_\rho^{U(1)}$ must be reduction mod 2. This is reasonable from the physical point of view, since it is known that a vortex in the $p+ip$ superfluid binds an odd number of Majorana zero modes, which carry the gravitational anomaly associated with the generator of $\Omega^2_\Spin$. We also learn that the map $f$ above sends the generator to the sum of the generators $(1,1) \in \bZ \oplus \bZ/2$, which is a bit more surprising! We will verify both these facts directly from the definition of these maps.

First we study $f$. In terms of spacetime manifolds, we want to take a 3-manifold $X$ with spin structure $\xi$ and a map $\phi \colon X \to S^1$, and construct a Spin$^c$ structure $A$ on $X$ under which $\phi$ has charge 2, so that $A$ gets Higgs'd to a spin structure. In terms of equations we want
\[2A = d\phi \\
dA = \pi w_2(TX) = \pi d\xi,\]
which can be solved by
\[A = \pi \xi + \frac{1}{2} d\phi.\]
The two terms here is the essential reason why we get the sum of generators when we compute $f$. It means when $\phi$ has an odd winding number around a 1-cycle of $X$, we twist the spin structure $\xi$ along that cycle, turning it from periodic to antiperiodic or vice versa.

We do the same thing when we compute ${\rm Ind}_\rho^{U(1)}$ according to the recipe given at the beginning of this subsection. There, from a spin surface $Y$ we form the manifold $X = Y \times S^1$ with $\phi$ winding once around the $S^1$ factor. The spin structure along this $S^1$ factor becomes twisted. When we evaluate the $\bZ$ generator of $\Omega^3_\Spin$ on this spin 3-manifold, we get the Arf invariant of $Y$ and its spin structure, which is the nontrivial element of $\Omega^2_\Spin$.

\subsection{Completing the circle}\label{subseccompletingthecircle}
By now we have defined our three maps: the residual family anomaly ${\rm Res}_\rho$, the defect anomaly ${\rm Def}_\rho$, and the index map ${\rm Ind}_\rho$. We have seen how they fit together into an exact sequence: the kernel of ${\rm Res}_\rho$ is the image of ${\rm Def}_\rho$ and the kernel of ${\rm Def}_\rho$ is the image of ${\rm Ind}_\rho$. In this section we will complete the circle and argue they form a \emph{long} exact sequence, in particular, the kernel of ${\rm Ind}_\rho$ is the image of ${\rm Res}_\rho$, so that this piece is exact:
\[\Omega_{G,s,\eta}^{D} \xrightarrow{{\rm Res}_\rho} \Omega^D_{G,s,\eta}(S(\rho))  \xrightarrow{{\rm Ind}_\rho} \Omega_{G,s,\eta+\rho}^{D-k+1}.\]

To see this, suppose we start with a class $\omega \in \Omega^D_{G,s,\eta}$ which we think of as a $G$-symmetric invertible theory. When we break the $G$ symmetry we get an equivariant invertible family over the ball $B(\rho)$ inside $V_\rho$. Restricting this family to $S(\rho)$ gives ${\rm Res}_\rho \omega$. Suppose we then compactify this family to form the invertible phase ${\rm Ind}_\rho {\rm Res}_\rho \omega$ and study it on a manifold $Y$ with boundary. If we place the family instead on the associated $B(\rho)$ bundle over $\partial Y$, then we get a trivial boundary condition of this invertible phase by gluing the two boundaries, which are both the associated $S(\rho)$ bundle over $\partial Y$. Compare Fig. \ref{fig:index-map}. Alternatively, the $\rho$-defect in the invertible phase $\omega$ is trivial, so it has trivial anomaly ${\rm Ind}_\rho {\rm Res}_\rho \omega = 0$. The converse follows from the Thom isomorphism.

\section{Examples}\label{sec:examples-of-SBLES}

In this section we collect a couple longer segments of the SBLES, containing some of the examples of individual maps we have already seen. 
Many more such examples can be found in \cite[\S 7, also see \S 8.3]{MathSmith}.

\subsection{$U(1)$ symmetry breaking for fermions}\label{subsubsecU1SBLES}

Let us consider the symmetry breaking long exact sequence for a $U(1)$ symmetry in a fermionic theory and an order parameter transforming in the charge 1 representation $\rho$. There are two cases to consider, depending on whether we have a spin-charge relation, meaning that fermionic operators have half-integer $U(1)$ charge, or not. In either case the relevant groups of invertible field theories we will need are shown in Table \ref{tab:U(1)_anomaly_groups}. To calculate these groups, one applies the universal property of Anderson duality, which is explained in \cite{MathSmith} to the spin bordism groups, the $\Spin\times\U(1)$ bordism groups, and the $\Spinc$ bordism groups, which are known: for spin bordism, see Milnor~\cite{Mil63}, for $\Omega_*^\Spin(B\U(1))$, see Wan-Wang~\cite[\S 3.1.5]{WW19}, and for \spinc bordism, see Bahri-Gilkey~\cite{BG87a}.

\begin{table}[htbp]
\centering
\setlength{\tabcolsep}{0.7em} 
\renewcommand{\arraystretch}{1.5} 
    \begin{tabular}{ C|C|C|C } 
     D & \Omega^D_\Spin & \Omega^D_{U(1),\Spin} & \Omega^D_{U(1),\Spin,\rho} = \Omega^D_{\Spin^c} \\ 
     \hline
     -1 & \Z & \Z & \Z\\ 
     0 & 0 & 0 & 0  \\ 
     1 & \Z/2 & \Z/2 \oplus \Z & \Z  \\
     2 & \Z/2 & \Z/2 & 0  \\ 
     3 & \Z & \Z^2 & \Z^2    \\
     4 & 0 & 0 & 0  \\ 
     5 & 0 & \Z^2 & \Z^2  \\ 
     6 & 0 & 0 & 0 \\
    \end{tabular}
    \caption{Classification of $D$-spacetime-dimensional fermionic invertible field theories  with $\bZ/2^F$, $U(1) \times \bZ/2^F$, and $U(1)^F$ symmetry, respectively.}
    \label{tab:U(1)_anomaly_groups}
\end{table}

First we study the case with spin-charge relation, where fermions carry half-charge under $U(1)$ and bosons carry integer charge. We consider symmetry breaking by a charge 1 order parameter (charge $2e$ from the point of view of the fermions). We studied such an example in \cref{subsubsecpipsuperfluid}, the $p+ip$ superfluid.

We organize the SBLES into rows associated with this symmetry breaking in each dimension $D$. The map from the first column to the second is the defect anomaly map ${\rm Def}_\rho$, from the second to the third is the residual family anomaly ${\rm Res}_\rho$, and the index maps ${\rm Ind}_\rho$ go from the third column of one row to the first column of the next. We omit arrows for maps that are zero, but the whole long exact sequence is connected. We use the isomorphism $\Omega^D_{U(1),\Spin,\rho}(S(\rho))  = \Omega^D_{\Spin}$ (see \eqref{eq:guage-to-pure-gravity-anomaly}) to substitute the latter group for the third map in the SBLES.

\begin{center}
\begin{tikzcd} [column sep=0.3in, row sep=small] 
	D & \Omega^{D-2}_{U(1),\Spin} & 
  \Omega^D_{\Spin^c} 
 & 
 \Omega^D_{\Spin}
 \\
	{-1} & 0 & \Z & \Z \\
	0 & 0 & 0 & 0 \\
	1 & \Z & \Z & {\Z/2} \\
	2 & 0 & 0 & {\Z/2} \\
	3 & {\Z/2\oplus\Z} & {\Z^2} & \Z \\
	4 & {\Z/2} & 0 & 0 \\
	5 & {\Z^2} & {\Z^2} & 0
 \arrow["{\rm Def}_\rho",from=1-2, to=1-3]
 \arrow["{\rm Res}_\rho",from=1-3, to=1-4]
	\arrow[from=2-3, to=2-4]
	\arrow[from=4-2, to=4-3]
	\arrow[from=4-3, to=4-4]
	\arrow[from=5-4, to=6-2, out=-20, in=160]
	\arrow[from=6-2, to=6-3]
	\arrow[from=6-3, to=6-4]
	\arrow[from=6-4, to=7-2, out=-20, in=160]
	\arrow[from=8-2, to=8-3]
\end{tikzcd}
    \label{fig:U(1)_LES_2}
\end{center}

The long subsequence beginning in $D = 2$ is
\begin{center}
\resizebox{1\hsize}{!}{
\begin{tikzcd}[ampersand replacement=\&, row sep= normal, scale = 0.6]\label{eqnU1subsequence}
	{\Omega^2_\Spin \cong \bZ/2} \& {\Omega^1_{U(1),\Spin} \cong \bZ/2 \oplus \bZ} \& {\Omega^3_{\Spin^c} \cong \bZ^2 } \\
	{\Omega^3_\Spin \cong \bZ} \& {\Omega^2_\Spin \cong \bZ/2} \&
	\arrow[from=1-1, to=1-2,"{\begin{psmallmatrix} 1 \\ 0 \end{psmallmatrix}}"]
	\arrow[from=1-2, to=1-3,"{\begin{psmallmatrix} 0 & 0 \\ 0 & 1 \end{psmallmatrix}}"]
	\arrow[from=1-3, to=2-1,"{\begin{psmallmatrix} 2 & 0 \end{psmallmatrix}}",out=-160,in=30]
	\arrow[from=2-1, to=2-2,"1"]
\end{tikzcd}}
\end{center}
We have discussed the last map in \cref{subsubsecpipsuperfluid}: it corresponds to the Majorana zero mode bound to the vortex of the $p+ip$ superfluid. Let us briefly discuss the computation of the other maps, although they are determined by the exact sequence.

The preceding map $\Omega^3_{\Spin^c} \to \Omega^3_\Spin$ measures the residual gravitational anomaly upon breaking the $U(1)$ symmetry. The group $\Omega^3_{\Spin^c}$ represents Chern-Simons terms associated with the four-dimensional invariants
\[ k_1 \left(\frac{1}{8} c_1^2 - \frac{1}{24} p_1\right) + k_2 c_1^2,\]
see \eqref{eqn2dspincanom}. Meanwhile the generator of $\Omega^3_\Spin$ is represented by $-\frac{1}{48} p_1$, so we see the map sends $(k_1,k_2)$ to $2k_1$.

The defect anomaly map $\Omega^1_{U(1),\Spin} \to \Omega^3_{\Spin^c}$ tells us the fermion parity as well as the $U(1)_\rho$ charge of the $\rho$-defect, i.e. the vortex of the order parameter. A physical model with anomaly $k_1 = 0$ and $k_2 = 1$ is the 1+1D compact boson with $U(1)$ acting only on the left movers. The vortex clearly is parity-even since there are no fermions in the model. However, it carries unit $U(1)_\rho$ charge, as is well-known from the chiral anomaly.

Finally, the index map $\Omega^2_\Spin \to \Omega^1_{U(1),\Spin}$ can be understood in terms of the ``topological superfluid'' in 1+1D. This can be thought of as a $U(1)$-charged Dirac fermion with the $U(1)$ symmetry broken by the two Majorana masses, which form a doublet. This is in the same phase as the Kitaev chain. A vortex operator in this phase, which changes the winding number of the order parameter, also changes the boundary conditions for the fermions, and therefore toggles the fermion parity of the ground state. This is captured by the nonzero index map, landing on the generator of $\Omega^1_\Spin \cong \bZ/2$ inside $\Omega^1_{U(1),\Spin}$, which gives the ``anomaly'' of the vortex operator, namely its fermion parity (compare \cref{subsubsecexamplethouless}).

Next we collect the SBLES for charge 1 breaking of a $U(1)$ symmetry without spin-charge relation:

    \begin{center}
    \begin{tikzcd}[ampersand replacement=\&, column sep=0.3in, row sep=small]
	D \& 
  \Omega^{D-2}_{\Spin^c} 
 \& \Omega^D_{U(1),\Spin} 
 \&
  \Omega^D_{\Spin} 
 \\
	{-1} \& 0 \& \Z \& \Z \\
	0 \& 0 \& 0 \& 0 \\
	1 \& \Z \& {\Z\oplus \Z/2} \& {\Z/2} \\
	2 \& 0 \& \Z/2 \& \Z/2 \\
	3 \& \Z \& {\Z^2} \& {\Z} \\
	4 \& 0 \& 0 \& 0 \\
	5 \& {\Z^2} \& {\Z^2} \& 0 \\
\arrow["{\rm Def}_\rho",from=1-2, to=1-3]
 \arrow["{\rm Res}_\rho",from=1-3, to=1-4]
	\arrow[from=2-3, to=2-4]
	\arrow[from=4-2, to=4-3]
	\arrow[from=4-3, to=4-4]
	\arrow[from=5-3, to=5-4]
	\arrow[from=6-2, to=6-3]
	\arrow[from=6-3, to=6-4]
	\arrow[from=8-2, to=8-3]
\end{tikzcd}
    
    \label{fig:U(1)_LES}
\end{center}

We have studied the map in $D = 5$ in \cref{defect_match_Dirac} when we considered breaking of chiral symmetry of a 4+1D Dirac fermion by its Dirac mass terms.

One general observation is that the index map always vanishes. The reason is that in the definition of the index map from \cref{subsecindexmap}, we produce an $S^1$ bundle $W$ with spin structure which extends to the disc bundle, since this $S^1$ always carries anti-periodic spin structure. Moreover, ${\rm Def}_\rho$ is an isomorphism from $\Omega^{D-2}_{U(1),\Spin,\rho}$ to the ``reduced'' part of $\Omega^D_{U(1),\Spin}$, namely those $U(1)$ symmetric invertible phases with no pure gravitational response, in other words which become trivial upon breaking the $U(1)$ symmetry. This the ``Smith isomorphism'' in \cite{KT90} (which can be proven following the methods of \cite{HKT19}, which considers the $\Z/2$ case).

\subsection{$\Z/2$ symmetry breaking for bosons}
\label{subsubsecZ2bosons}

Now let us discuss perhaps the simplest example of the SBLES, which describes the breaking of a unitary $\bZ/2$
symmetry of a bosonic system by a single real order parameter transforming in the sign representation $\sigma$. 
On
the domain wall, this unitary symmetry is transmuted to an anti-unitary symmetry. For reference, the relevant
classification groups are shown in Table \ref{tab:Z2_anomaly_groups_bosonic}, with $\Omega^D_{SO}$ denoting
$D$-spacetime-dimensional bosonic invertible field theories, $\Omega^D_{\bZ/2,SO}$ denoting those with a unitary $\bZ/2$ symmetry, and $\Omega^D_{\bZ/2,SO,\sigma}$ denoting those with an anti-unitary $\bZ/2$ symmetry. As usual, these groups were obtained by applying Anderson duality to oriented bordism, unoriented bordism, and the oriented bordism of $B\Z/2$. See Thom~\cite[Théorèmes IV.9, IV.13]{ThomThesis} for oriented and unoriented bordism groups in low degrees.
We do not know of an explicit reference for $\Omega_*^\SO(B\Z/2)$, but it can be calculated using a result of Wall~\cite{Wal60} that implies that the Atiyah-Hirzebruch spectral sequence for oriented bordism collapses for any space whose mod $p$ cohomology vanishes for all odd $p$.

\begin{table}[htbp]
\centering
\setlength{\tabcolsep}{0.7em}
\renewcommand{\arraystretch}{1.5} 
    \begin{tabular}{ C|C|C|C } 
     D & \Omega^D_{SO} & \Omega^D_{\bZ/2,SO,\sigma} = \Omega^D_{O} & \Omega^D_{\bZ/2,SO} \\
     \hline
     -1 & \Z & 0 & \Z \\
     0 & 0 & \Z/2 & 0 \\
     1 & 0 & 0 & \Z/2 \\
     2 & 0 & \Z/2 & 0 \\
     3 & \Z & 0 & \Z\oplus \Z/2 \\
     4 & 0 & (\Z/2)^2 & 0 \\
     5 & \Z/2 & \Z/2 & (\Z/2)^3\\
     6 & 0 & (\Z/2)^3 & \Z/2 \\
     7 & 0 & \Z/2 & (\Z/2)^3 \\
    \end{tabular}
    \caption{Classification of $D$-spacetime-dimensional bosonic invertible field theories with no symmetry, time reversal symmetry, and $\bZ/2$ symmetry respectively.}
    \label{tab:Z2_anomaly_groups_bosonic}
    \end{table}

We collect the SBLES as follows. By \eqref{eq:guage-to-pure-gravity-anomaly} that the third group in the SBLES simplifies:
$\Omega^D_{\bZ/2,SO}(S(\sigma)) = 
 \Omega^D_{SO}$.
    \begin{center}
\begin{tikzcd}[column sep=0.3in]
	& 
 \Omega^{D-1}_O & \Omega^D_{\bZ/2,SO} &  
 \Omega^D_{SO} \\
	{-1} & 0 & \Z & \Z \\
	0 & 0 & 0 & 0 \\
	1 & {\Z/2} & {\Z/2} & 0 \\
	2 & 0 & 0 & 0 \\
	3 & {\Z/2} & {\Z/2\oplus\Z} & \Z \\
	4 & 0 & 0 & 0 \\
	5 & {(\Z/2)^2} & {(\Z/2)^3} & {\Z/2} \\
	6 & {\Z/2} & {\Z/2} & 0 \\
	7 & {(\Z/2)^3} & {(\Z/2)^3} & 0
\arrow["{\rm Def}_\sigma",from=1-2, to=1-3]
 \arrow["{\rm Res}_\sigma",from=1-3, to=1-4]
 \arrow[from=4-2, to=4-3]
	\arrow[from=2-3, to=2-4]
	\arrow[from=6-2, to=6-3]
	\arrow[from=6-3, to=6-4]
	\arrow[from=8-2, to=8-3]
	\arrow[from=8-3, to=8-4]
	\arrow[from=9-2, to=9-3]
	\arrow[from=10-2, to=10-3]
 
\end{tikzcd}
    \label{fig:Z2_bosonic_LES}
\end{center}

This has a similar structure to the $U(1) \times \bZ/2^F \to \bZ/2^F$ breaking we studied above in \cref{subsubsecU1SBLES}, splitting into isomorphisms given by ${\rm Def}_\sigma$ (the ``Smith isomorphism'') and ${\rm Res}_\sigma$, with ${\rm Ind}_\sigma$ vanishing.
Meanwhile, the pure gravitational part is mapped isomorphically by ${\rm Res}_\sigma$, since by definition we do not need the $\bZ/2$ symmetry to detect it, and $\bZ/2$ acts transitively on $S(\sigma)$, so the residual family anomaly is determined by the anomaly of the unbroken subgroup, which is just the gravitational part.

We can also compute the SBLES associated with breaking of a time reversal symmetry by a single real order parameter transforming in the sign representation. This turns out to be more interesting, since we no longer have a Smith isomorphism, and ${\rm Ind}_\sigma$ may be nonzero.
There is a similar identification of the group over the sphere as $\Omega^D_{\bZ/2,SO,\sigma}(S(\sigma)) = \Omega^D_{SO}$.
\begin{center}
\begin{tikzcd}[column sep=0.3in, row sep=small]
	& \Omega^{D-1}_{\bZ/2,SO} & 
 \Omega^{D}_O & 
 \Omega^D_{SO}\\
	{-1} & 0 & 0 & \Z \\
	0 & \Z & {\Z/2} & 0 \\
	1 & 0 & 0 & 0 \\
	2 & {\Z/2} & {\Z/2} & 0 \\
	3 & 0 & 0 & \Z \\
	4 & {\Z\oplus \Z/2} & {\Z/2 \oplus \Z/2} & 0 \\
	5 & 0 & {\Z/2} & {\Z/2} \\
	6 & {(\Z/2)^3} & {(\Z/2)^3} & 0 \\
	7 & {\Z/2} & {\Z/2} & 0
 \arrow["{\rm Def}_\sigma",from=1-2, to=1-3]
 \arrow["{\rm Res}_\sigma",from=1-3, to=1-4]
	\arrow[from=2-4, to=3-2, out=-20, in=160]
	\arrow[from=3-2, to=3-3]
	\arrow[from=5-2, to=5-3]
	\arrow[from=6-4, to=7-2, out=-20, in=160]
	\arrow[from=7-2, to=7-3]
	\arrow[from=8-3, to=8-4]
	\arrow[from=9-2, to=9-3]
	\arrow[from=10-2, to=10-3]
\end{tikzcd}
    \label{fig:Z2_bosonic_LES_2}
\end{center}

Consider for example the 3rd to 4th rows. We have the sequence
\begin{center}
    \begin{tikzcd}[ampersand replacement=\&, row sep=normal]
    {\Omega^3_{\bZ/2,SO,\sigma}(S(\sigma))} \ar[r] \ar[d,phantom,"{\cong}" {rotate=-90}] \& {\Omega^3_{\bZ/2,SO}} \ar[r] \ar[d,phantom,"{\cong}" {rotate=-90}] \& {\Omega^4_O} \ar[d,phantom,"{\cong}" {rotate=-90}] \\
    \Z \ar[r,"{\begin{psmallmatrix} 2 & 0 \end{psmallmatrix}}"] \& \Z\oplus \Z/2 \ar[r,"{\begin{psmallmatrix} 1 & 0 \\ 0 & 1 \end{psmallmatrix}}"] \& \Z/2\oplus \Z/2
    \end{tikzcd}
\end{center}

The generator of the first nonzero group is the $S(\sigma) \cong S^0$-family with an $E_8$ phase \cite{Lu_2012} at one point (the generator of $\Omega^3_{SO} \cong \bZ$), and its inverse phase at the other point. To compute the index map, we study a domain wall between the $E_8$ and its inverse, which with the standard boundary conditions has chiral modes with $c_L = 16$, $c_R = 0$. The induced unitary $\bZ/2$ symmetry is anomaly-free, since $k = 0$ mod 8 of the modes are charged. This theory represents the anomaly $(2,0) \in \bZ \oplus \bZ/2 \cong \Omega^3_{\bZ/2,SO}$.

The next map sends the $E_8$ state, representing $(1,0)$ in that group, to the time-reversal symmetric phase described by a gravitational $\theta = \pi$ angle, or $\frac{1}{2} w_2^2$. This encodes the well-known fact that the time reversal domain wall at the boundary of that theory (known as $e_f m_f$ in \cite{Wang_2014}) hosts $c_L = 8$ mod 16 gapless chiral modes. Meanwhile, it sends the Levin-Gu SPT \cite{Levin_2012} associated to $\frac{1}{2} A^3$ and representing $(0,1)$ in $\Omega^3_{\bZ/2,SO}$, to the phase associated with $\frac{1}{2} w_1^4$.

\subsection{$\bZ/2$ symmetry breaking for fermions}
\label{subsubsecZ2fermions}

Now we turn to the same $\bZ/2$ symmetry breaking scenario for fermions.
In the fermionic setting, there are four different types of $\bZ/2$ symmetry:
either unitary with $U^2 = 1$ or $U^2 = (-1)^F$, or time reversing with $T^2 = 1$ or $T^2 = (-1)^F$, as discussed in \cref{subsecanomalyintro}. The relevant classifications are collection in Table \ref{tab:Z2_anomaly_groups_fermionic}, corresponding to low-degree bordism groups that are explicitly calculated in the following references: Milnor~\cite{Mil63} (spin bordism),
Giambalvo~\cite{Gia73} ($\Pin^+$ bordism),
Kirby-Taylor~\cite{KT90} ($\Pin^-$ bordism),
García-Etxebarria and Montero~\cite[(C.18)]{GEM18} ($\Spin\times\Z/2$ bordism),\footnote{This calculation, or more precisely its equivalent analogue in $\ko$-homology, was first done by Mahowald-Milgram~\cite{MM76}, with $\ko_*(B\Z/2)$ worked out explicitly by Bruner-Greenlees~\cite[Example 7.3.1]{BG10}, but the cited reference lists spin bordism groups specifically.}
and Giambalvo~\cite{Gia73a} ($\Spin\times_{\Z/2}\Z/4$ bordism).

\begin{widetext}
\begin{table*}[htbp]
\centering
\setlength{\tabcolsep}{0.7em}
\renewcommand{\arraystretch}{1.5} 
    \begin{tabular}{ C|C|C|C|C|C } 
     D & \Omega^D_\Spin & \Omega^D_{\bZ/2,\Spin} & \Omega^D_{\bZ/2,\Spin,\sigma} = \Omega^D_{{\rm Pin}^-} & \Omega^D_{\bZ/2,\Spin,2\sigma} = \Omega^D_{\Spin \times_{\bZ/2} \bZ/4} & \Omega^D_{\bZ/2,\Spin,3\sigma} = \Omega^D_{{\rm Pin}^+} \\ 
     \hline
     -1 & \Z & \Z & 0 & \Z & 0 \\ 
     0 & 0 & 0 & \Z/2 & 0 & \Z/2 \\ 
     1 & \Z/2 & (\Z/2)^2 & \Z/2 & \Z/4 & 0 \\
     2 & \Z/2 & (\Z/2)^2 & \Z/8 & 0 & \Z/2 \\ 
     3 & \Z & \Z \oplus \Z/8 & 0 & \Z & \Z/2 \\
     4 & 0 & 0 & 0 & 0 & \Z/{16} \\ 
     5 & 0 & 0 & 0 & \Z/{16} & 0 \\ 
     6 & 0 & 0 & \Z/{16} & 0 & 0 \\
    \end{tabular}
    \caption{Fermionic invertible field theories in $D$ spacetime dimensions with symmetry $\bZ/2^F$, $\bZ/2^U \times \bZ/2^F$, $\bZ/2^T \times \bZ/2^F$, $\bZ/4^U$, or $\bZ/4^T$, respectively.}
    \label{tab:Z2_anomaly_groups_fermionic}
    \end{table*}
    \end{widetext}

There are four different SBLES concerning symmetry breaking by the order parameter $\sigma$, one for each of the four types of $\bZ/2$ symmetry. 
We have computed an initial segment of each. Observe that by \eqref{eq:guage-to-pure-gravity-anomaly}, we have $\Omega^D_{\bZ/2,\Spin,n\sigma}(S(\sigma)) = \Omega^D_{\Spin}$ for any $n$, so the third group in the long exact sequence classifies pure gravitation anomalies.
First we study $\bZ/2^U \times \bZ/2^F$ breaking to $\bZ/2^F$. 

\vspace{-3 pt}
\begin{center}
\begin{tikzcd}[column sep=0.3in, row sep=small]
	& \Omega^{D-1}_{{\rm Pin}^-} & \Omega^D_{\bZ/2,\Spin} & 
 \Omega^D_\Spin \\
	{-1} & 0 & \Z & \Z \\
	0 & 0 & 0 & 0 \\
	1 & {\Z/2} & {(\Z/2)^2} & {\Z/2} \\
	2 & {\Z/2} & {(\Z/2)^2} & {\Z/2} \\
	3 & {\Z/8} & {\Z/8\oplus \Z} & \Z \\
	4 & 0 & 0 & 0 \\
	5 & 0 & 0 & 0 \\
	6 & 0 & 0 & 0 \\
	7 & {\Z/16} & {\Z/16\oplus \Z^2} & {\Z^2}
\arrow["{\rm Def}_\sigma",from=1-2, to=1-3]
 \arrow["{\rm Res}_\sigma",from=1-3, to=1-4]
	\arrow[from=2-3, to=2-4]
	\arrow[from=4-2, to=4-3]
	\arrow[from=4-3, to=4-4]
	\arrow[from=5-2, to=5-3]
	\arrow[from=5-3, to=5-4]
	\arrow[from=6-2, to=6-3]
	\arrow[from=6-3, to=6-4]
	\arrow[from=10-2, to=10-3]
	\arrow[from=10-3, to=10-4]
\end{tikzcd}
\end{center}

Next we study $\bZ/2^T \times \bZ/2^F$ breaking to $\bZ/2^F$. 
\begin{center}
\begin{tikzcd}[column sep=0.3in, row sep=small]
	D & \Omega^{D-1}_{\bZ/2,\Spin, 2\sigma} & 
 \Omega^D_{\Pinm} & 
 \Omega^D_{\Spin} \\
	{-1} & 0 & 0 & \Z \\
	0 & \Z & {\Z/2} & 0 \\
	1 & 0 & {\Z/2} & {\Z/2} \\
	2 & {\Z/4} & {\Z/8} & {\Z/2} \\
	3 & 0 & 0 & \Z \\
	4 & \Z & 0 & 0 \\
	5 & 0 & 0 & 0 \\
	6 & {\Z/16} & {\Z/16} & 0 \\
	7 & 0 & 0 & {\Z^2}
 \arrow["{\rm Def}_\sigma",from=1-2, to=1-3]
 \arrow["{\rm Res}_\sigma",from=1-3, to=1-4]
	\arrow[from=2-4, to=3-2, out=-20, in=160]
	\arrow[from=3-2, to=3-3]
	\arrow[from=4-3, to=4-4]
	\arrow[from=5-2, to=5-3]
	\arrow[from=5-3, to=5-4]
	\arrow[from=6-4, to=7-2, out=-20, in=160]
	\arrow[from=9-2, to=9-3]
\end{tikzcd}
\end{center}
One generator of $\Omega^2_\Pinm \cong \bZ/8$ is represented by a $T$-odd Majorana zero mode. Upon forgetting the $T$ symmetry, this still has a gravitational anomaly, associated with $\Omega^2_\Spin \cong \bZ/2$. If we have two $T$-odd Majoranas $\gamma_{1,2}$, we can write the $T$-odd pairing $i\gamma_1 \gamma_2$ which leads to a unique ground state. Changing the sign of this term toggles the fermion parity of this ground state, so the associated operator has unit charge under the induced unitary symmetry $U$, since $U^2 = (-1)^F$. This ``anomaly'' represents a generator of $\Omega^1_{\bZ/2,\Spin,2\sigma} \cong \bZ/4$.

Next we have the breaking of a unitary symmetry $U$ with $U^2 = (-1)^F$ down to $\bZ/2^F$  as well as breaking of a time reversal symmetry $T$ with $T^2 = (-1)^F$ down to $\bZ/2^F$. 
\begin{center}
\begin{tikzcd}[column sep=0.3in, row sep=small]
	D & 
 \Omega^{D-1}_{\Pinp} & \Omega^D_{\bZ/2,\Spin,2\sigma} & 
 \Omega^D_{\Spin} \\
	{-1} & 0 & \Z & \Z \\
	0 & 0 & 0 & 0 \\
	1 & {\Z/2} & {\Z/4} & {\Z/2} \\
	2 & 0 & 0 & {\Z/2} \\
	3 & {\Z/2} & \Z & \Z \\
	4 & {\Z/2} & 0 & 0 \\
	5 & {\Z/16} & {\Z/16} & 0 \\
 \arrow["{\rm Def}_\sigma",from=1-2, to=1-3]
 \arrow["{\rm Res}_\sigma",from=1-3, to=1-4]
	\arrow[from=8-2, to=8-3]
	\arrow[from=5-4, to=6-2, out=-20, in=160]
	\arrow[from=6-3, to=6-4]
	\arrow[from=6-4, to=7-2, out=-20, in=160]
	\arrow[from=2-3, to=2-4]
	\arrow[from=4-2, to=4-3]
	\arrow[from=4-3, to=4-4]
\end{tikzcd}
\end{center}
\noindent\rule[0.3ex]{\linewidth}{1pt}
\begin{center}
\begin{tikzcd}[column sep=0.3in, row sep=small]
	D & \Omega^{D-1}_{\bZ/2,\Spin} & 
 \Omega^D_{\Pinp} & 
 \Omega^D_{\Spin} \\
	{-1} & 0 & 0 & \Z \\
	0 & \Z & {\Z/2} & 0 \\
	1 & 0 & 0 & {\Z/2} \\
	2 & {(\Z/2)^2} & {\Z/2} & {\Z/2} \\
	3 & {(\Z/2)^2} & {\Z/2} & \Z \\
	4 & {\Z\oplus \Z/8} & {\Z/16} & 0 \\
    \arrow["{\rm Def}_\sigma",from=1-2, to=1-3]
    \arrow["{\rm Res}_\sigma",from=1-3, to=1-4]
	\arrow[from=2-4, to=3-2, out=-20, in=160]
	\arrow[from=3-2, to=3-3]
	\arrow[from=4-4, to=5-2, out=-20, in=160]
	\arrow[from=5-2, to=5-3]
	\arrow[from=5-4, to=6-2, out=-20, in=160]
	\arrow[from=6-2, to=6-3]
	\arrow[from=6-4, to=7-2, out=-20, in=160]
	\arrow[from=7-2, to=7-3]
\end{tikzcd}
\end{center}
The short exact sequence from $D = 3$ to $D = 4$ was analyzed in \cref{subsubsectdwformajorana} in the context of time reversal domain walls of $2+1$D Majorana fermions.

Finally, we may also consider composing two of the defect maps in this section, which produces a defect map and an SBLES corresponding to symmetry breaking by an order parameter transforming under $2\sigma$. Note that an $2\sigma$ order parameter is a pair of $\Z/2$-odd operators.
Unlike symmetry breaking by an $\sigma$ order parameter, 
this process preserves the (anti)-unitarity of the symmetry operator, and exchanges the symmetry types $\Z/2^T\times \Z/2^F$ and $\Z/4^T$, and symmetry types $\Z/2^U\times\Z/2^F$ and $\Z/4^U$. 
We discuss the case of $\Z/4^T$ ($\Pinp$) breaking to $\Z/2^T\times \Z/2^F$ ($\Pinm$). 
Note that the residual family anomaly is more than just gravitational, as it falls outside of the cases considered in \cref{eq:guage-to-pure-gravity-anomaly}. In particular, it is classified by
$\Omega^D_{\Z/2, \Spin, \sigma}(S(2\sigma)) = \tilde{\Omega}^{D+1}_{\Spin}(\mathbb{R}P^2)$.

\begin{center}
\begin{tikzcd}[column sep=0.34in]
	& {\Omega^{D-2}_{\mathrm{Pin}^-}} & {\Omega^{D}_{\mathrm{Pin}^+}} & {\tilde{\Omega}^{D+1}_{\Spin}(\mathbb{R}P^2)} \\
	{-1} & 0 & 0 & 0 \\
	0 & 0 & {\Z/2} & {\Z/2} \\
	1 & 0 & 0 & {\Z/2} \\
	2 & {\Z/2} & {\Z/2} & {\Z/4} \\
	3 & {\Z/2} & {\Z/2} & {\Z/2} \\
	4 & {\Z/8} & {\Z/16} & {\Z/2} \\
	5 & 0 & 0 & 0 \\
	6 & 0 & 0 & 0
	\arrow[from=3-3, to=3-4]
	\arrow[from=4-4, to=5-2, out=-20, in=160]
	\arrow[from=5-3, to=5-4]
	\arrow[from=5-4, to=6-2, out=-20, in=160]
	\arrow[from=6-3, to=6-4]
	\arrow[from=7-2, to=7-3]
	\arrow[from=7-3, to=7-4]
	\arrow[from=1-2, to=1-3, "\mathrm{Def}_{2\sigma}"]
	\arrow[from=1-3, to=1-4, "\mathrm{Res}_{2\sigma}"]
\end{tikzcd}
\end{center}
Consider the $D=4$ short exact sequence from the table above. As noted above, the generator of $\Z/16$ can be represented by $2+1$D Majorana fermions. One Majorana fermion has a single $T$-odd Majorana mass term $\cO_1$, and we have considered its symmetry breaking. However, as discussed in \cref{subsubsec2dMajoranaresidualanomaly}, we can't find a second $T$-odd operator such that 
\begin{equation}\label{eq:this-eq}
    r \cos \theta \cO_1 + r \sin \theta \cO_2
\end{equation}
is nondegenerately gapped for large $r$ and all $\theta$. This corresponds to the $\Z/2$ residual anomaly in $\Omega^4_{\Spin}(S(2\sigma))$. However, if we have two Majorana fermions, then we can find two $T$-odd operators such that 
\eqref{eq:this-eq} gaps the system for all $\theta$. The codimensional two vortex is precisely the Majorana zero modes, which have a $\Z/8$ classification.

\subsection{$\bZ/3$ symmetry breaking for fermions}
\label{subsubsecZ3fermions}

\begin{table}[htbp]
\centering
\setlength{\tabcolsep}{0.7em}
\renewcommand{\arraystretch}{1.5} 
    \begin{tabular}{L|C|C}
        D & \Omega^D_{\Spin} & \Omega^D_{\bZ/3,\Spin} \\
        \hline 
        -1 & \Z  & \Z \\
        0 & 0  & 0 \\
        1 & \Z/2  & \Z/3\oplus \Z/3 \\
        2 & \Z/2 &  \Z/2 \\
        3 & \Z  & \Z\oplus \Z/3 \\
        4 & 0  & 0 \\
        5 & 0  & \Z/9 \\
        6 & 0  & 0 \\
        7 & \Z^2  & \Z^2\oplus \Z/9 \\
        8 & 0  & 0 \\
        9 & (\Z/2)^2 &  (\Z/2)^2\oplus \Z/3\oplus \Z/{27} \\
        10 & (\Z/2)^3 &  (\Z/2)^3 \\
        11 & \Z^3 &  \Z^3\oplus \Z/3\oplus \Z/{27} \\
    \end{tabular}
    \caption{Classification of invertible field theories with $\bZ/2^F$ and $\bZ/3 \times \bZ/2^F$ symmetry in $D$ spacetime dimensions.}
        \label{tab:Z3classification}
    \end{table}

An interesting case which demonstrates some of the more typical complexity of the SBLES is $\bZ/3$ symmetry breaking in fermionic systems via a charge 1 order parameter. Such a symmetry must be unitary and the symmetry group must have the product structure $\bZ/3^U \times \bZ/2^F$.
The relevant classification is shown in Table \ref{tab:Z3classification}; the new piece of information we need is $\Omega_*^\Spin(B\Z/3)$, worked out in degrees $11$ and below in~\cite[\S 12.2]{DDHM22} using work of Bruner-Greenlees~\cite[Example 7.3.2]{BG10}.

The anomaly group over the sphere $S(\rho)$ simplifies as $\Omega^D_{\bZ/3,\Spin}(S(\rho)) = \Omega^D_{\Spin} \oplus \Omega^{D-1}_{\Spin}$, and 
the long exact sequence is the following:
\onecolumngrid
\begin{center}
    \begin{tikzcd} 
    [ampersand replacement=\&, column sep=0.3in, ]
        D \& \Omega^{D-2}_{\bZ/3,\Spin} \& \Omega^D_{\bZ/3,\Spin} \& 
        \Omega^D_{\Spin} \oplus \Omega^{D-1}_{\Spin} \\
        {-1} \& 0 \& \Z \& \Z \\
        0 \& 0 \& 0 \& {\Z} \\
        1 \& \Z \& {\Z/2\oplus \Z/3} \& {\Z/2} \\
        2 \& 0 \& {\Z/2} \& {(\Z/2)^2} \\
        3 \& {\Z/2\oplus\Z/3} \& {\Z\oplus \Z/3} \& {\Z\oplus \Z/2} \\
        4 \& {\Z/2} \& 0 \& \Z \\
        5 \& \Z\oplus \Z/3 \& \Z/9 \& 0 \\
        6 \& 0 \& 0 \& 0 \\
        7 \& \Z/9 \& \Z^2\oplus\Z/9 \& \Z^2 \\
        8 \& 0 \& 0 \& \Z/2 \\
        9 \& \Z/2\oplus \Z/9 \& {(\Z/2)^2\oplus \Z/3\oplus\Z/27} \& {(\Z/2)^2} \\
        10 \& 0 \& (\Z/2)^3 \& (\Z/2)^5 \\
        11 \& {(\Z/2)^2\oplus \Z/3\oplus \Z/27} \& {\Z^3\oplus \Z/3\oplus \Z/27} \& \Z^3\oplus(\Z/2)^3
        \arrow["{\rm Def}_\rho",from=1-2, to=1-3]
 \arrow["{\rm Res}_\rho",from=1-3, to=1-4]
        \arrow["(1)",from=2-3, to=2-4]
    \arrow["(3)"{description}, from=3-4, to=4-2, out=-30, in=150]
        \arrow["{\binom{0}{1}}",from=4-2, to=4-3]
        \arrow["{(1,0)}",from=4-3, to=4-4]
        \arrow[from=5-3, to=5-4]
        \arrow["{\tiny\begin{pmatrix} 1 & 0 \\ 0 & 0\end{pmatrix}\normalsize}"{description},from=5-4, to=6-2, out=-30, in=150]
        \arrow["{\tiny\begin{pmatrix} 0 & 0 \\ 0 & 1\end{pmatrix}\normalsize}",from=6-2, to=6-3]
        \arrow["{\tiny\begin{pmatrix} 1 & 0 \\ 0 & 0\end{pmatrix}\normalsize}"', from=6-3, to=6-4]
    \arrow["{(0,1)}"{description},from=6-4, to=7-2, out=-30, in=150]
    \arrow["{\tiny\begin{pmatrix} -3 \\ 1 \end{pmatrix}\normalsize}"{description},from=7-4, to=8-2, out=-30, in=150]
        \arrow["{(1,3)}",from=8-2, to=8-3]
        \arrow["{(0,1)}",from=10-2, to=10-3]
        \arrow["{\binom{1}{0}}",from=10-3, to=10-4]
    \arrow[from=11-4, to=12-2, out=-30, in=150]
        \arrow[from=12-2, to=12-3]
        \arrow[from=12-3, to=12-4]
        \arrow[from=13-3, to=13-4]
    \arrow[from=13-4, to=14-2, out=-30, in=150]
        \arrow[from=14-2, to=14-3]
        \arrow[from=14-3, to=14-4]
    \end{tikzcd}
    \end{center}
\twocolumngrid
    Note that because there is no twist, $\Omega^D_{\bZ/3,\Spin} = \tilde \Omega^D_{\bZ/3,\Spin} \oplus \Omega^D_{\Spin}$, where $\tilde \Omega^D_{\bZ/3,\Spin}$ denotes the subgroup of those phases which become trivial upon breaking $\bZ/3$. This subgroup is finite and has no 2-torsion, so ${\rm Res}_\rho$ is always zero on it, while it maps the $\Omega^D_{\Spin}$ factor injectively. It follows that the long exact sequence splits into a series of short exact sequences of the form
    \begin{equation}
        0 \to \Omega^{D-2}_{\Spin} \xrightarrow{{\rm Ind}_\rho} \Omega^{D-2}_{\bZ/3,\Spin} \xrightarrow{{\rm Def}_\rho} \tilde \Omega^D_{\bZ/3,\Spin} \to 0 
    \end{equation}
    There are four interesting ones:
    \begin{itemize}
        \item $D = 1$: $\mb{Z} \to \mb{Z} \to \mb{Z}/3$
        \item $D = 2$: $\mb{Z}/2 \to \mb{Z}/2 \oplus \mb{Z}/3 \to \mb{Z}/3$.
        \item $D = 4$: $\mb{Z}/2 \to \mb{Z}/2$.
        \item $D = 5$: $\mb{Z} \to \mb{Z} \oplus \mb{Z}/3 \to \mb{Z}/9$. 
    \end{itemize}
    Let us consider for example $D = 5$. The first $\bZ \cong \Omega^3_{\Spin} = \tilde \Omega^4_{\bZ/3,\Spin}(S(\rho))$ is generated by a 3+1D family which pumps a generator of $\Omega^3_{\Spin}$ to the boundary over each third of the $S(\rho) \cong S^1$. When we compute the first map, the index map, we look at the vortex where the order parameter windings all the way around $S(\rho)$. This has three 1+1D gapless Majorana modes of the same chirality, with $\bZ/3$ acting as a permutation. This can be written as a neutral chiral Majorana and a charge 1 Weyl, so it has anomaly $(3,1) \in \bZ \oplus \bZ/3$. The calculation of the next map, the defect anomaly map, follows \cref{defect_match_Dirac}.

$D = 1$ is also interesting. Since it involves phases in ``negative dimension'' we need to think in terms of families (compare \cref{subsubsecberryphaseprojrep}). The map ${\rm Def}_\rho \colon\bZ \to \bZ/3$ says that if we have an $S^2$ family of quantum states, with $\bZ/3$ acting as a $2\pi/3$ polar rotation, the difference in the $\bZ/3$ charges of the states at the poles equals the Chern number mod 3.

\subsection{$\bZ/4$ symmetry breaking for fermions}
\label{subsubsecZ4fermions}
Now we consider symmetry breaking of a unitary symmetry $U$ with $U^4 = (-1)^F$ by a charge 1 order parameter (defining the representation $\rho$).
The relevant classifications are given in Table \ref{tab:Z4classification}; the new bordism groups we need as input are $\Omega_*^\Spin(B\Z/4)$ and $\Omega_*^{\Spin\times_{\Z/2}\Z/8}$, which appear explicitly in~\cite[\S 12.1, \S 13.2]{DDHM22} (the former building on a calculation of Bruner-Greenlees~\cite[Example 7.3.3]{BG10}).
Finally, there is an isomorphism $\Omega^D_{\bZ/4,\Spin,\rho}(S(\rho)) = \Omega^D_{\Spin} \oplus \Omega^{D-1}_{\Spin}$.

\begin{table}[htbp]
\centering
\setlength{\tabcolsep}{0.7em}
\renewcommand{\arraystretch}{1.5} 
    \begin{tabular}{ C|C|C|C }     
     D & \Omega^D_{\Spin} &  \Omega^D_{\bZ/4,\Spin} & \Omega^D_{\bZ/4,\Spin,\rho} \\ 
     \hline
     -1 & \Z  & \Z & \Z \\ 
     0 & 0  & 0 & 0 \\ 
     1 & \Z/2   & \Z/2 \oplus \Z/4 & \Z/8 \\
     2 & \Z/2   & (\Z/2)^2 & 0 \\ 
     3 & \Z &  \Z\oplus \Z/2\oplus \Z/8 & \Z\oplus \Z/2 \\
     4 & 0  & 0 & 0   \\ 
     5 & 0  & \Z/4 & \Z/{32}\oplus \Z/2  \\ 
     6 & 0  & 0 & 0 \\
     
    \end{tabular}
    \caption{The classification of $\bZ/4$ symmetric invertible field theories in $D$ spacetime dimension. Here $\rho$ is the charge one representation of $\bZ/4$, giving a unitary symmetry class with $U^4 = (-1)^F$.}
    \label{tab:Z4classification}
    \end{table}

    The symmetry breaking long exact sequence is as follows:
    \begin{center}
    \begin{tikzcd}
    [ampersand replacement=\&, column sep=0.1in, row sep=0.3in]
        D \& \Omega^{D-2}_{\bZ/4,\Spin} \& \Omega^D_{\bZ/4,\Spin,\rho}\& 
        \Omega^D_{\Spin} \oplus \Omega^{D-1}_{\Spin} \\
        {-1} \& 0 \& \Z \& \Z \\
        0 \& 0 \& 0 \& {\Z} \\
        1 \& \Z \& {\Z/8} \& {\Z/2} \\
        2 \& 0 \& 0 \& {(\Z/2)^2} \\
        3 \& {\Z/2\oplus\Z/4} \& {\Z\oplus \Z/2} \& {\Z\oplus \Z/2} \\
        4 \& {(\Z/2)^2} \& 0 \& \Z \\
        5 \& \Z\oplus \Z/8 \oplus \Z/2 \& \Z/32\oplus\Z/2 \& {0} \\
        6 \& 0 \& 0 \& 0 
        \arrow["{\rm Def}_\rho",from=1-2, to=1-3]
 \arrow["{\rm Res}_\rho",from=1-3, to=1-4]
        \arrow["(1)", from=2-3, to=2-4]
    \arrow["(4)"{description}, from=3-4, to=4-2, out=-30, in=150]
        \arrow[from=4-2, to=4-3]
        \arrow[from=4-3, to=4-4]
    \arrow["{\tiny\begin{pmatrix} 1 & 0 \\ 0 & 2\end{pmatrix}\normalsize}"{description},from=5-4, to=6-2, out=-30, in=150]
        \arrow["{\tiny\begin{pmatrix} 0 & 0 \\ 0 & 1\end{pmatrix}\normalsize}",from=6-2, to=6-3]
        \arrow["{\tiny\begin{pmatrix} 2 & 0 \\ 0 & 0\end{pmatrix}\normalsize}"', from=6-3, to=6-4]
    \arrow["{\tiny\begin{pmatrix} 1 & 0 \\ 0 & 1\end{pmatrix}\normalsize}"{description},from=6-4, to=7-2, out=-30, in=150]
    \arrow[from=7-4, to=8-2, out=-30, in=150]
        \arrow[from=8-2, to=8-3]
    \end{tikzcd}
    \end{center}

    Let us study the subsequence from $D = 2$ to $D = 4$.
    The first map is ${\rm Ind}_\rho \colon \Omega^2_{\bZ/4,\Spin,\rho}(S(\rho)) \to \Omega^1_{\bZ/4,\Spin}$. The $\Omega^2_{\Spin}$ generator is the 1+1D topological superfluid we discussed around \eqref{eqnU1subsequence} test and gets mapped to the $\Omega^1_{\Spin}$ generator as we discussed there. The other $\bZ/2$ generator pumps four fermionic charges to the boundary when traversing $S(\rho) \cong S^1$. Let $\cO_i$, $i = 1,2,3,4$ be the four operators creating these charges, which anticommute. The $\bZ/4$ symmetry acts on them by $\cO_i \mapsto \cO_{i+1}$. The vortex operator of the whole family is the product $\cO_1 \cO_2 \cO_3 \cO_4$, which we compute is charge 2 under $\bZ/4$. This corresponds to $2 \in \bZ/4 \cong \tilde \Omega^1_{\bZ/4,\Spin}$.

    The next group is $\Omega^3_{\bZ/4,\Spin,\rho} \cong \bZ \oplus \bZ/2$. The $\bZ$ generator represents the anomaly of a charge $1/2$ (charge 1 under $\bZ/8^F$) 1+1D Weyl fermion, while the $\bZ/2$ generator represents that of a Dirac fermion with chiral charges $\pm 1/2$ for the left and right handed components. In the second case, if we break the symmetry by adding a Dirac mass (which transforms in the representation $\rho$) we get a Thouless pump with a unit $\bZ/4$-charged vortex operator, matching the defect anomaly map $\bZ/4 \to \bZ/2$. ${\rm Res}_\rho$ maps the $\bZ$ generator to two times the $\bZ$ generator of $\Omega^3_{\Spin}$, since a Weyl fermion is two Majorana-Weyl fermions.

    Another interesting subsequence goes from $D = 4$ to 5, in particular exactness requires the index map to be 
    \begin{center}
        \begin{tikzcd}[ampersand replacement=\&]
            {\rm Ind}_\rho\colon \Omega^4_{\bZ/4,\Spin}(S(\rho)) \ar[r
            ] \ar[d,phantom,"{\cong}" {rotate=-90}] \& \Omega^3_{\bZ/4,\Spin} \ar[d,phantom,"{\cong}" {rotate=-90}] \\
            \bZ \ar[r,"{\left( 4 \quad  1 \quad  0 \right)^T}"] \& \bZ \oplus \bZ/8 \oplus \bZ/2.
        \end{tikzcd}
    \end{center}
    Let us verify this. The generator of the source is a family which pumps the generator of $\Omega^3_{\Spin} \cong \bZ$ to the boundary over each quarter of the circle $S(\rho)$. When we form the $\rho$-defect, we have four copropagating 1+1D chiral Majorana modes, with $\bZ/4$ acting as a permutation. This corresponds to a charge 1 and a charge 2 left-handed Weyl. If this was a $U(1)$ symmetry, its chiral anomaly would be $1^2 + 2^2 = 5$, which is indeed coprime to 8, so when $U(1)$ is reduced to $\bZ/4$, this is a generator of $\bZ/8$.

\subsection{$SU(2)$ symmetry breaking for fermions}
\label{subsubsecSU2fermions}

Now we discuss $SU(2)$ and $SO(3)$ symmetry breaking in fermion systems. There are three cases of interest, $SU(2) \times \bZ/2^F$, $SO(3) \times \bZ/2^F$, and $SU(2)^F$, where the latter has a spin-charge relation where fermions carry half integer spin and bosons carry integer spin. We will consider symmetry breaking by both spin-$1/2$ and spin-1 order parameters. The relevant classifications are shown in Table \ref{tab:SU2_anomaly_groups}. 
Once again we refer the reader to \cite[\S 7]{MathSmith} for the computation of the anomaly groups and long exact sequences.
As input, we need $\Omega_*^\Spin$, as discussed above, and several families of bordism groups that have not yet appeared in this paper.
\begin{itemize}
    \item $\Omega_*^\Spin(B\SO(3))$ is calculated in low degrees by Wan-Wang~\cite[\S 5.3.3]{WW19}.
    \item $\Omega_*^\Spin(B\SU(2))$ is calculated in low degrees by Lee-Tachikawa~\cite[Appendix B.2]{LT21}.
    \item $\Omega_*^{\Spin^h}$ is calculated in low degrees by Freed-Hopkins~\cite[Theorem 9.97]{FH16}.
\end{itemize}

\begin{table}[htbp] 
\centering
\setlength{\tabcolsep}{0.7em}
    \renewcommand{\arraystretch}{1.5} 
    \begin{tabular}{ C|C|C|C|C }
     D & \Omega^D_{\Spin} & \Omega^D_{SU(2), \Spin} & \Omega^D_{SO(3),\Spin} & \Omega^D_{SO(3),\Spin,\mathbf{1}}\\ 
     \hline
     -1 & \Z & \Z & \Z &\Z \\ 
     0 & 0 & 0 & 0 & 0\\ 
     1 & \Z/2 & \Z/2  &\Z/2 & 0\\
     2 & \Z/2 & \Z/2 &(\Z/2)^2 &0\\ 
     3 & \Z & \Z^2  & \Z^2 &\Z^2\\
     4 & 0 & 0  & 0 &0\\ 
     5 & 0 & \Z/2 & 0 &(\Z/2)^2\\ 
     6 & 0 & \Z/2 & \Z/2 &(\Z/2)^2\\
     7 & \Z^2 & \Z^4& \Z^4 & \Z^4\\
    \end{tabular}
\caption{Anomaly groups relevant to the $SU(2)$ families of long exact sequences of field theories}
\label{tab:SU2_anomaly_groups}
\end{table}

First we will consider $SU(2) \times \bZ/2^F$ symmetry breaking to $\bZ/2^F$ by a complex spin-$1/2$ order parameter, which is the simplest case.
We use that $\Omega^D_{SU(2),\Spin}(S(\rho)) = 
 \Omega^D_{\Spin}$ (see \eqref{eq:guage-to-pure-gravity-anomaly}). Note that this is another instance of the ``Smith isomorphism'' where the long exact sequence splits and $\Omega^{D-4}_{SU(2),\Spin} \simeq \tilde{\Omega}^D_{SU(2)}(\Spin)$.
\begin{center}
\begin{tikzcd}[row sep=small, column sep = 0.3in]
	D & \Omega^{D-4}_{SU(2),\Spin} & \Omega^D_{SU(2),\Spin} & 
 \Omega^D_{\Spin} \\
	{-1} & 0 & \Z & \Z \\
	0 & 0 & 0 & 0 \\
	1 & 0 & \Z/2 & \Z/2\\
	2 & 0 & \Z/2 & \Z/2 \\
	3 & \Z & {\Z^2} & {\Z} \\
	4 & 0 & 0 & 0 \\
	5 & \Z/2 & \Z/2 & 0 \\
	6 & {\Z/2} & \Z/2 & 0 \\
	7 & \Z^2 & {\Z^4} & {\Z^2}
 \arrow["{\rm Def}_\rho",from=1-2, to=1-3]
 \arrow["{\rm Res}_\rho",from=1-3, to=1-4]
	\arrow[from=2-3, to=2-4]
	\arrow[from=4-3, to=4-4]
	\arrow[from=5-3, to=5-4]
	\arrow[from=6-2, to=6-3]
	\arrow[from=6-3, to=6-4]
  \arrow[from=8-2, to=8-3]
  \arrow[from=9-2, to=9-3]
  \arrow[from=10-2, to=10-3]
  \arrow[from=10-3, to=10-4]
\end{tikzcd}
\end{center}
The generator of $\Omega^5_{SU(2),\Spin} \cong \bZ/2$ corresponds to Witten's $SU(2)$ anomaly \cite{witten19822}. For example, we can consider $N_f = 2$ QCD with chiral $SU(2)_L \times SU(2)_R$ symmetry. In the usual chiral symmetry breaking scenario, the order parameters are mass terms and form a complex $SU(2)$ doublet. The defect anomaly map here is capturing the fact that skyrmions in this theory are fermions.

Next we study $SU(2) \times \bZ/2^F$ symmetry breaking to $U(1) \times \bZ/2^F$ by a real spin-$1$ order parameter.
Note that
$ \Omega^D_{SU(2),\Spin}(S(\rho)) = 
 \Omega^D_{U(1),\Spin}$ as $SU(2)$ acts surjectively on $S(\rho)$ with stabilizer group $U(1)$ (see \eqref{eq:guage-to-pure-gravity-anomaly}).
\begin{center}
\begin{tikzcd}[row sep=small, column sep = small]
	D & \Omega^{D-3}_{SU(2),\Spin} & \Omega^D_{SU(2),\Spin} & 
 \Omega^D_{U(1),\Spin} \\
	{-1} & 0 & \Z & \Z \\
	0 & 0 & 0 & 0 \\
	1 & 0 & \Z/2 & \Z/2 \oplus \bZ\\
	2 & \Z & \Z/2 & \Z/2 \\
	3 & 0 & {\Z^2} & {\Z^2} \\
	4 & \Z/2 & 0 & 0 \\
	5 & \Z/2 & \Z/2 & \Z^2 \\
	6 & {\Z^2} & \Z/2 & 0 \\
	7 & 0 & {\Z^4} & {\Z^4}
 \arrow["{\rm Def}_\rho",from=1-2, to=1-3]
 \arrow["{\rm Res}_\rho",from=1-3, to=1-4]
	\arrow[from=2-3, to=2-4]
	\arrow[from=4-3, to=4-4]
	\arrow[from=5-3, to=5-4]
	\arrow[from=4-4, to=5-2,out=-20,in=160]
 \arrow[from=6-4, to=7-2,out=-20,in=160]
 \arrow[from=8-4, to=9-2,out=-20,in=160]
	\arrow[from=6-3, to=6-4]
  \arrow[from=8-2, to=8-3]
  \arrow[from=9-2, to=9-3]
  \arrow[from=10-3, to=10-4]
\end{tikzcd}
\end{center}

The residual family anomaly in $D = 3$ maps the gravitation Chern-Simons term associated with $\Omega^3_{\Spin} \cong \bZ$ to itself, while the level 1 $SU(2)$ Chern-Simons term corresponding to the other generator of $\Omega^3_{SU(2),\Spin}$ maps to a level 2 Chern-Simons term for the unbroken $U(1)$ subgroup. If we have a level 1 Chern-Simons term, the $\rho$-defect acts as a $U(1)$ monopole (this is like an 't Hooft-Polyakov monopole), and is thus fermionic, which is captured by the index map.

Now we study $SO(3) \times \bZ/2^F$ symmetry breaking to $U(1) \times \bZ/2^F$ by a real spin 1 order parameter.
Once again we use that $\Omega^D_{SO(3),\Spin}(S(\rho)) = 
 \Omega^D_{U(1),\Spin}$.
\begin{center}
\begin{tikzcd}[row sep=small, column sep = small]
	D & \Omega^{D-3}_{SO(3),\Spin,\mathbf{1}} & \Omega^D_{SO(3),\Spin} & 
 \Omega^D_{U(1),\Spin} \\
	{-1} & 0 & \Z & \Z \\
	0 & 0 & 0 & 0 \\
	1 & 0 & {\Z/2} & {\Z\oplus\Z/2} \\
	2 & \Z & {(\Z/2)^{\oplus 2}} & {\Z/2} \\
	3 & 0 & {\Z^2} & {\Z^2} \\
	4 & 0 & 0 & 0 \\
	5 & 0 & 0 & {\Z^2} \\
	6 & {\Z^2} & {\Z/2} & 0 \\
	7 & 0 & {\Z^4} & {\Z^4}
 \arrow["{\rm Def}_\rho",from=1-2, to=1-3]
 \arrow["{\rm Res}_\rho",from=1-3, to=1-4]
	\arrow["(1)", from=2-3, to=2-4]
	\arrow["{(0, 1)}", from=4-3, to=4-4]
	\arrow["{(2, 0)}"{description}, from=4-4, to=5-2, out=-20, in=160]
	\arrow["{(1, 0)}"', from=5-2, to=5-3]
	\arrow["{(0, 1)}", from=5-3, to=5-4]
	\arrow["\cong", from=6-3, to=6-4]
	\arrow["{(1, 2)}"{description}, from=8-4, to=9-2, out=-20, in=160]
	\arrow["{(1, 0)}"', from=9-2, to=9-3]
	\arrow[from=10-3, to=10-4]
\end{tikzcd}
\end{center}

Finally we study $SU(2)^F$ symmetry breaking to $U(1)^F$ by a real spin 1 order parameter.
Note that $\Omega^D_{SO(3),\Spin}(S(\rho)) \cong \Omega^D_{\Spin^c}$.
\begin{center}
\begin{tikzcd}[row sep=small, column sep = small]
	D & \Omega^{D-3}_{SO(3),\Spin} & \Omega^D_{SO(3),\Spin,\mathbf{1}} & 
 \Omega^D_{\Spin^c} \\
	{-1} & 0 & \Z & \Z \\
	0 & 0 & 0 & 0 \\
	1 & 0 & 0 & \Z \\
	2 & \Z & 0 & 0 \\
	3 & 0 & {\Z^2} & {\Z^2} \\
	4 & {\Z/2} & 0 & 0 \\
	5 & {(\Z/2)^{\oplus 2}} & {(\Z/2)^{\oplus 2}} & {\Z^2} \\
	6 & {\Z^2} & {(\Z/2)^{\oplus 2}} & 0 \\
	7 & 0 & {\Z^4} & {\Z^4}
	\arrow["(1)", from=2-3, to=2-4]
	\arrow["(1)"{description}, from=4-4, to=5-2, out=-20, in=160]
 \arrow["{\rm Def}_\rho",from=1-2, to=1-3]
 \arrow["{\rm Res}_\rho",from=1-3, to=1-4]
	\arrow["{(2, 1)}", from=6-3, to=6-4]
	\arrow["{(1, 0)}"{description}, from=6-4, to=7-2, out=-20, in=160]
	\arrow["\cong", from=8-2, to=8-3]
	\arrow["2"{description}, from=8-4, to=9-2, out=-20, in=160]
	\arrow["{\bmod 2}"', from=9-2, to=9-3]
	\arrow[from=10-3, to=10-4]
\end{tikzcd}
\end{center}

\section{The symmetry breaking long exact sequence in group cohomology}\label{sec:group_cohomology}

So far, we have been assuming the SPT-cobordism conjecture \cref{sss:g_anom}. However, our symmetry breaking long exact sequence exists for any classification, including the group cohomology classification \cite{Chen_2013}. In this setting, we have explicit formulas for the three maps. Indeed, the SBLES is equivalent to the Thom-Gysin sequence for the fibration
\[S(\rho) \to S(\rho)//G \xrightarrow{\pi} BG.\]
The space in the middle is the \emph{homotopy quotient}
\[S(\rho)//G \coloneqq EG \times_G S(\rho),\]
where $EG$ is a contractible space with a free $G$ action. Recall that $BG$ is defined as $EG/G$. By projecting onto the $EG$ coordinate then, we define a map
\[\pi\colon S(\rho)//G \to BG.\]
We can think of this as the $S(\rho)$ bundle over $BG$ associated to the universal $G$ bundle. The importance of $S(\rho)//G$ is that its ordinary cohomology equals the (Borel) $G$-equivariant cohomology of $S(\rho)\colon$
\[H^D(S(\rho)//G,A) = H^D_G(S(\rho),A),\]
where $A$ is some arbitrary coefficients. Given a representation $\eta$ of $G$, we get a $U(1)$ local system over $S(\rho)//G$ by pulling back the usual $U(1)$ local system over $BG$ (twisted by the determinant line of $\eta$). This defines
\[H^D(S(\rho)//G,U(1)^\eta) = H^D_G(S(\rho),U(1)^\eta),\]
which classifies group cohomology invariants of anomaly-free $G$-equivariant invertible families, where $\eta$ remembers which symmetries of $G$ are anti-unitary. In other words, these are $\eta$-twisted $U(1)$-valued $D$-forms
$\zeta(A,\phi)$
on spacetime, constructed from a $G$ gauge field $A$ and a section $\phi$ of the $S(\rho)$ bundle associated to the $G$ gauge bundle.

The Thom-Gysin sequence gives us the following formulas
\begin{equation}
    \begin{aligned}
        {\rm Def}_\rho \colon H^{D-k}(BG,U(1)^{\eta + \rho}) &\to H^D(BG,U(1)^\eta) \\
    \alpha &\mapsto e(\rho) \cup \alpha,
    \end{aligned}
\end{equation}
where
$e(\rho) \in H^k(BG,\bZ^\rho)$
is the Euler class of $\rho$.
\begin{equation}
    \begin{aligned}
        {\rm Res}_\rho : H^D(BG,U(1)^\eta) &\to H^D(S(\rho)//G,U(1)^\eta) \\
\omega &\mapsto \pi^* \omega,
    \end{aligned}
\end{equation}
where $\pi^*$ is the pullback along the projection above. Finally
\begin{equation}
\begin{aligned}
    {\rm Ind}_\rho \colon H^D(S(\rho)//G,U(1)^\eta) &\to H^{D-k+1}(BG,U(1)^{\eta+\rho}) \\
\zeta &\mapsto \int_{S(\rho)} \zeta,
\end{aligned}
\end{equation}
where this integral indicates integration against the ($\rho$-twisted) homology class of the $S(\rho)$ fiber in $S(\rho)//G$.

\section{Outlook}\label{sec:outlook}

In this paper, we have presented a long exact sequence in symmetry breaking, relating three maps: the residual family anomaly which captures the equivariant family anomaly when we move around the order parameter space and which gives the obstruction to having a local $\rho$-defect, the defect anomaly map which reconstructs the bulk anomaly from that of the $\rho$-defect, and the index map which describes the anomaly of the $\rho$-defect in an anomaly-free equivariant family on a sphere and describes how the different symmetry breaking patterns are distinguished by their $\rho$-defects. The kernel of each map is the image of the next, connecting anomaly matching formulas for a given group and representation in all dimensions.

There are a few directions for future work we think are promising. The first is to better understand how to formulate the twisted symmetry $G_\rho$ on the lattice. We can use the CPT symmetry to obtain this symmetry in Lorentz invariant, unitary theories, as described in \cite{HKT19}. However, on the lattice there may be no such symmetry and it is not clear how to proceed.

One approach which seems fruitful is to make contact with the recent anomaly approaches to Lieb-Schultz-Mattis (LSM) theorems \cite{Po_2017,Yang_2018,Else_2020}. In particular, we can think of the reconstruction of the bulk $G_\rho$ anomaly from the $G_\rho$ anomaly of the $\rho$-defect as a pure point-group LSM theorem. Indeed, in this case the LSM map of \cite{Else_2020} (see Appendix I there) is given by cup with the Euler class of $\rho$ and agrees with the defect anomaly map we computed. It seems that the two anomalies are related by the crystalline equivalence principle, which we intend to revisit in future work.

The SBLES is a convenient tool for computing classifications of anomalies in different symmetry classes, since different symmetry breaking patterns can be combined to obtain more constraints on the classification group in terms of lower dimensional groups, and the maps are often determined by exactness. This approach is complementary to the ``decorated domain wall'' methods \cite{Chen_2014}, which are mathematically formalized as an Atiyah-Hirzebruch spectral sequence \cite{Gaiotto_2019,wang2021domain,Thorngren_2021,shiozaki2023generalized}. In these methods, low dimensional invertible phases are glued together to form higher dimensional ones, allowing one to bootstrap the classification, simply knowing the gluing rules. These rules however, known as the spectral sequence differentials, have still not been completely computed. However, the physical interpretation of these differentials (see for instance \cite{shiozaki2023generalized}) matches the index map we have defined, and it seems possible that all differentials may be computable in terms of it. This is a direction we are currently exploring.

Another interesting direction is to see if the residual family anomalies we have defined yield new results for quantum field theories. Indeed, we know of no other proof even for the statement in Section \ref{subsubsec2dMajoranaresidualanomaly} regarding two $T$-odd deformations of the 2+1D free Majorana fermion. This may yield new predictions on the phase diagrams near such models.

The $\rho$-defect is a very symmetric defect, and vortices and other defects which occur in experiments will not have this symmetry, especially in problems involving vortex and skyrmion scattering or lattices. It would be very interesting to understand how such symmetry breaking imprints on the anomalous modes at the vortex core. When $G$ acts projectively in the core, presumably the degeneracy is split when the vortex is not symmetric, but we expect there will be a non-trivial Berry phase in this splitting to match the anomaly.

\bibliography{bib}
 \bibliographystyle{alpha}

 \end{document}